\documentclass[11pt]{article}

\usepackage[preprint]{acl}

\usepackage{times}
\usepackage{latexsym}
\usepackage{enumitem}
\usepackage{xurl}

\usepackage[T1]{fontenc}
\usepackage[utf8]{inputenc}

\usepackage{microtype}

\usepackage{inconsolata}

\usepackage{tabularx, booktabs, longtable}

\usepackage{graphicx}
\usepackage[table]{xcolor}
\usepackage{multirow}
\usepackage{booktabs}
\usepackage{amssymb}
\usepackage{pifont}
\usepackage{makecell}
\newcommand{\cmark}{\ding{51}}
\newcommand{\xmark}{\ding{55}}

\usepackage{listings}
\usepackage{xcolor}
\usepackage{caption} % 用于控制 Listing 的标题格式

\usepackage{needspace}

\usepackage[most]{tcolorbox}

\newcommand{\prefix}{\textit{Pref}\textbf{Ix}}
\usepackage{pifont} % 用于提供 \ding{51} (对勾) 和 \ding{55} (叉号)
\usepackage{pifont}
\usepackage{xcolor} % 用于提供颜色支持
\usepackage{setspace}

\newtcolorbox{instructionbox}{
  colback=gray!5,
  boxrule=0pt,
  leftrule=3pt,
  colframe=gray!50,
  sharp corners,
  boxsep=5pt,
  top=5pt,
  bottom=5pt
}

\providecommand{\cmark}{}
\providecommand{\xmark}{}

\renewcommand{\cmark}{\textcolor{green!50!black}{\ding{51}}}
\renewcommand{\xmark}{\textcolor{red}{\ding{55}}}

\title{\textit{Pref}Ix: Understand and Adapt to User Preference\\in Human-Agent Interaction}
\author{
  Jialin Li$^{1}$ \quad
  Zhenhao Chen$^{2}$ \quad
  Hanjun Luo$^{1}$ \quad
  Hanan Salam$^{1}$ \\
  \\
  $^{1}$ New York University Abu Dhabi, Abu Dhabi, UAE \\
  $^{2}$ Mohamed bin Zayed University of Artificial Intelligence, Abu Dhabi, UAE
}

\begin{document}
\maketitle
\begin{abstract}
LLM-based agents can complete tasks correctly yet still frustrate users through poor interaction patterns, such as excessive confirmations, opaque reasoning, or misaligned pacing. Current benchmarks evaluate task accuracy but overlook how agents interact: whether they infer preferences from implicit cues, adapt dynamically, or maintain fine-grained interaction quality. We introduce \prefix, a configurable environment that evaluates both what agents accomplish and how they interact. Central to \prefix~is the Interaction-as-a-Tool (IaaT) paradigm, which treats interaction behaviors as structured tool calls, unifying them with existing evaluation frameworks. We define 31 preference settings across 14 attributes and formalize user experience (UX) as a core metric alongside task accuracy. A composite LLM-as-a-Judge mechanism across seven UX dimensions achieves strong aggregate reliability (ICC $>$ 0.79), high internal consistency ($\alpha$ = 0.943), and human correlation ($\rho$ = 0.52--0.78). Preference-aware agents show 7.6\% average UX improvement and 18.5\% gain in preference alignment. Our work is openly accessible \href{https://github.com/JL10897/ix_personalization_2.git}{\textit{here}}.
% Our work is openly accessible \href{https://anonymous.4open.science/r/ix_personalization_2-5FF6}{\textit{here}}.
\end{abstract}

\section{Introduction}

Adapting to human preferences has been a fundamental pursuit in modern AI systems \cite{QuPreference}. By aligning LLMs with the distinct needs and experiences of individuals or groups, the systems facilitate frictionless and engaging interactions, fostering improved user experiences.

Early work on personalization distinguishes \textit{content} (what facts or recommendations the model produces) from \textit{presentation} (how information is expressed through style or tone) \cite{ryan-etal-2025-synthesizeme}. However, this framing is constrained by the limited interaction space of language models. With the emergence of LLM-based agents, personalization can extend beyond generation to encompass richer interaction behaviors. 

% [DONE] Compress to 1–2 sentences, add agent bridge. Original feedback:
% - You over-elaborate subcategories; underspecify why this matters for agents.
% - Keep the distinction and one or two examples.
% - End with clear bridge: "this framing is constrained by limited interaction space..."

LLM-based agents build upon LLMs' language capabilities by adding planning, tool orchestration, and environment grounding, facilitating a higher degree of task automation \cite{li2025surveypersonalizationragagent}. These capabilities broaden the scope of personalization across the entire pipeline, from understanding user preferences, through planning and execution, to generation, introducing opportunities for behavioral adaptation that shape user experience even when task goals remain constant.

\begin{figure*}[ht]
    \centering
    \includegraphics[width=0.7\textwidth]{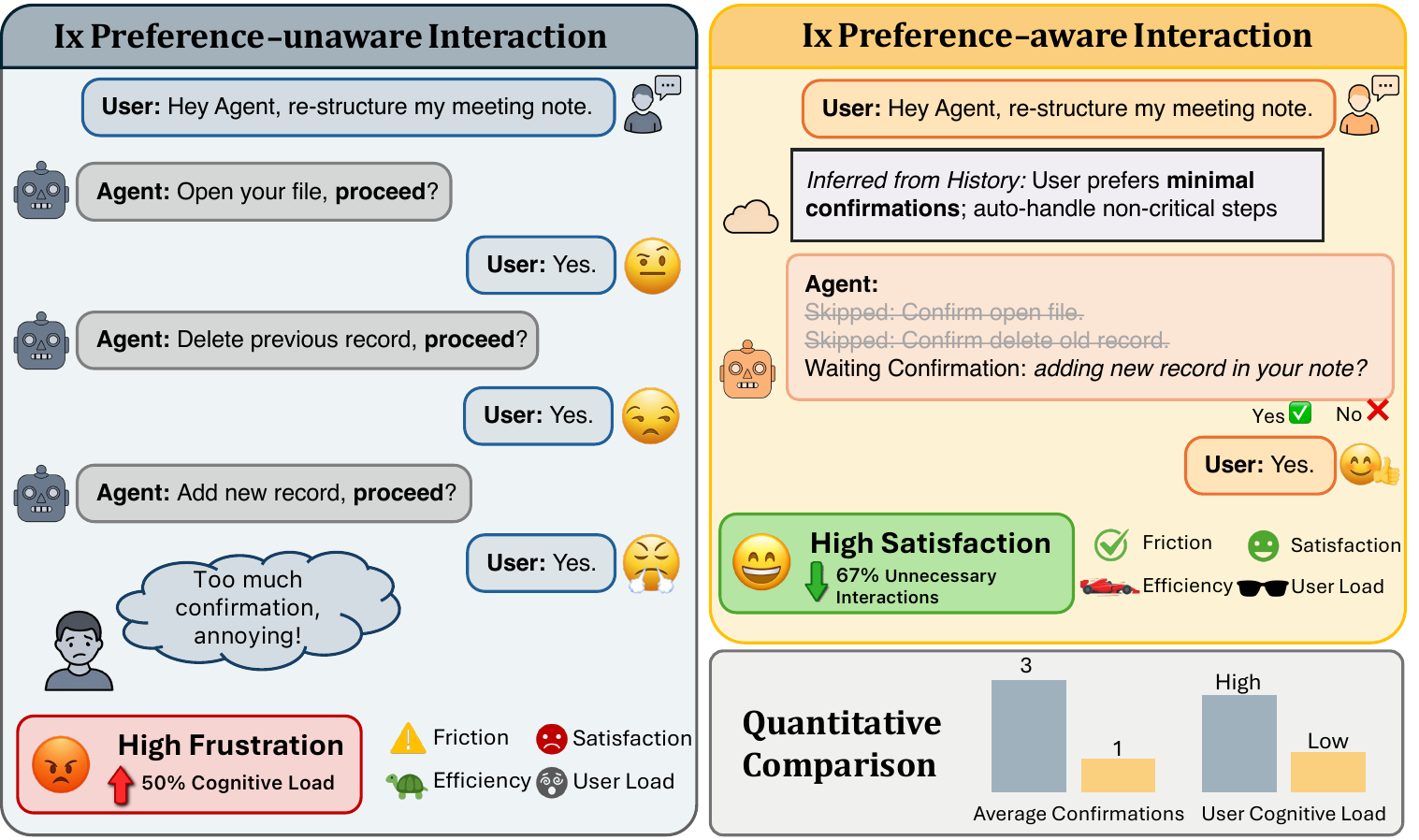}
    \caption{\textbf{Impact of Preference Awareness in Interaction.} By inferring user preferences from history, the agent reduces unnecessary turns (right), whereas a rigid agent (left) leads to user frustration despite correct task execution.}
    \vspace{-0.35cm}
    \label{fig:intuition}
\end{figure*}

To illustrate, Figure~\ref{fig:intuition} shows that the distinction lies not merely in task completion but in behavioral patterns during execution: while a rigid agent induces frustration through excessive confirmations, a preference-aware agent adapts its confirmation frequency and execution pacing to lower user load.

Beyond this example, language agents' interaction behaviors span multiple dimensions: \textit{transparency and auditability}, which captures how much visibility users want into tool parameters and information sources; \textit{interaction pace and flow}, which concerns how information is presented, collected, and sequenced; \textit{robustness and adaptability}, which addresses how agents handle uncertainty and failure; and \textit{strategy and initiative}, which determines how proactive the agent should be. An agent's capability in varying these choices based on their understanding of the user's interaction preference can drastically affect user experience (UX) even when task outcomes are identical. Consequently, evaluation must capture not only task correctness but also interaction-level UX.

% [DONE] Added preference categories and UX evaluation motivation.

Current agent benchmarks focus on task success \cite{huang2024planningcreationusagebenchmarking}, reasoning and planning \cite{gioacchini2024agentquest}, memory \cite{maharana2024evaluating}, tool use \cite{huang2024planningcreationusagebenchmarking}, robustness \cite{debenedetti2024agentdojo,yao2024tau}, and safety \cite{qiu2024evaluating,chen2024agentpoison,yuan2024r,zhang2024cybench}, largely measured via output correctness in controlled settings \cite{mohammadi2025evaluation}. They rarely model or evaluate how agents adapt their interaction behavior to user preferences.
% [DONE] Compressed evaluation survey; kept citations attached to each concept.

As the field moves toward human-centered agents, recent work has begun exploring adaptation to individual goals. For example, \citet{qian2025userbench} focused on agents' ability to resolve underspecified preferences in multi-turn interactions. However, these works adapt content to underspecified user goals but treat the interaction protocol (confirmation strategy, pacing, exposure of intermediate steps) as fixed.
% [DONE] Added concrete limitation of UserBench.

On the other hand, UX highly relies on users' interaction preferences \cite{augstein2023personalized}. Yet UX metrics have received limited attention in Natural Language Processing (NLP) evaluation, especially in human-agent interaction. UX metrics like satisfaction and efficiency are only occasionally introduced when they are related to token cost or response latency \cite{kang2025win}. Even if there is experience related evaluation, they are highly inconsistent and are often treated as supplementary analyses \cite{pettersson2024measurement} rather than core metrics. Therefore, despite widespread discussion, UX evaluation remains loose and heterogeneous, lacking unified, comparable, and reproducible formal representation.

Hence, we identify a fundamental evaluation gap in current agent benchmarks: the challenge is not whether to consider UX, but how to incorporate interaction-level UX as a core, reproducible, and scalable evaluation target, rather than treating it as auxiliary qualitative analysis. Following this pipeline, three gaps emerge: (1) at the \textit{understanding} layer, benchmarks rely on explicit preference specifications~\cite{salemi2024lamp,hao2025evaluating} rather than testing inference from implicit behavioral cues~\cite{li2025surveypersonalizationragagent}; (2) at the \textit{planning and execution} layer, trajectory-based metrics exist for static settings~\cite{chen2024teval,ma2024agentboard}, yet evaluation in dynamic user simulation with long-term adaptation remains underexplored~\cite{mohammadi2025evaluation}; (3) at the \textit{generation} layer, coarse-grained preference alignment is measured, but fine-grained interaction quality dimensions are lacking~\cite{wang2024understanding,li2025surveypersonalizationragagent}. The open problem is to design an evaluation framework where interaction-level UX, grounded in user preferences, is (i) operationalized as agent trajectories, (ii) measurable at scale, and (iii) comparable across models and tasks.

To address these gaps, we present \prefix{}, which integrates a simulation environment, an evaluation framework linking user preferences to agent behaviors, and validated UX metrics for automated assessment. Here is the breakdown of the contributions:
\vspace{-0.2cm}

\begin{itemize}[leftmargin=2em,itemsep=-0.1em]
% [DONE] Added framing sentence before contributions; defined IaaT; clarified taxonomy; unified terminology to "UX".

    \item[\ding{182}] \textbf{Unifying Interaction and Tool-use Framework:} With \prefix, we introduce a configurable interactive environment for multi-turn tool-use scenarios. By proposing the Interaction-as-a-Tool (IaaT) paradigm, where four categories of interaction behaviors (e.g., confirmation requests) are modeled as callable tools alongside external APIs, we unify the evaluation of interaction processes within existing agent frameworks.

    \item[\ding{183}] \textbf{Structured Evaluation of Task Correctness and UX:} We define a taxonomy of 14 interaction preference attributes across four categories (e.g., confirmation frequency, error handling style), and map them to observable agent behaviors. This places UX on equal footing with task accuracy: agents are judged both on what they accomplish and on how they conduct the interaction.

    \item[\ding{184}] \textbf{Reliable Automated Assessment:} We validate a composite multiple LLM-as-a-Judge framework across seven UX dimensions. Experiments demonstrate traceable, reproducible, reliable, and scalable assessment.

\end{itemize}

\section{Related work}
\begin{table*}[t]
    \centering
    \small
    \setlength{\tabcolsep}{3.8pt} % Reduce column padding
    \renewcommand{\arraystretch}{1} % 稍微增加一点行高防止拥挤
    \caption{Comparison of existing environments for human-agent interaction, comparing Multiturn support, Function Calling, User Simulation, Evaluation Method (Rule-based vs. Generative), Evaluation Target (Process vs. Interaction), and Interaction Preference handling.}
    \label{tab:bench_survey}
    \begin{tabular}{lcccccccc}
    \toprule
    \multirow{2}{*}{\textbf{Benchmark}} & 
    \multirow{2}{*}{\textbf{Multiturn}} & 
    \multirow{2}{*}{\makecell[c]{\textbf{Function}\\\textbf{Calling}}} & 
    \textbf{User Simulation} & 
    \multicolumn{2}{c}{\textbf{Eval Method}} & 
    \multicolumn{2}{c}{\textbf{Eval Target}} &
    \textbf{Preference} \\
    \cmidrule(lr){4-4} \cmidrule(lr){5-6} \cmidrule(lr){7-8} \cmidrule(lr){9-9}
    & & & Dynamic & Rule-based & Generative & Process & Interaction & Interaction \\
    \midrule
    $\tau$ Bench (\citeyear{yao2024tau}) & \cmark & \cmark & \cmark & \cmark & \xmark & \xmark & \xmark & \xmark \\
    IN3 (\citeyear{qian2024tell}) & \cmark & \xmark & \cmark & \cmark & \xmark & \cmark & \xmark & \xmark \\
    TravelPlanner (\citeyear{xie2024travelplanner}) & \xmark & \cmark & \xmark & \cmark & \xmark & \cmark & \xmark & \xmark \\
    BFCL v3 (\citeyear{patil2025bfcl}) & \cmark & \cmark & \xmark & \cmark & \xmark & \cmark & \xmark & \xmark \\
    ETAPP (\citeyear{hao2025evaluating}) & \xmark & \cmark & \xmark & \cmark & \cmark & \cmark & \cmark & \xmark \\
    UserBench (\citeyear{qian2025userbench}) & \cmark & \cmark & \cmark & \cmark & \xmark & \cmark & \cmark & \xmark \\
    \midrule
    \textbf{\textit{Pref}Ix (ours)} & \cmark & \cmark & \cmark & \cmark & \cmark & \cmark & \cmark & \cmark \\
    \bottomrule
    \end{tabular}
\end{table*}

\subsection{User-centered Evaluation Targets of Language Agents}

User-centric evaluation in human-agent interaction emerged from the need to assess not only task correctness but also appropriateness of the interaction process: whether agents interact efficiently and align with diverse user expectations \cite{ryan2025synthesizeme,qian2025userrl}. Recent work has advanced multi-turn evaluation~\cite{chakraborty2025t1}, dynamic user simulation~\cite{yao2024tau}, and user preference modeling~\cite{qian2025userbench,singh2024personal}. These gaps manifest across the personalization pipeline. At the \textit{understanding} layer, benchmarks such as LaMP~\cite{salemi2024lamp} evaluate explicit profile extraction, yet inference from implicit behavioral cues remains undertested~\cite{li2025surveypersonalizationragagent}. At the \textit{planning and execution} layer, trajectory-based metrics assess tool accuracy in static settings~\cite{chen2024teval,ma2024agentboard}, but lack evaluation in dynamic user simulation with long-term adaptive behavior~\cite{mohammadi2025evaluation}. At the \textit{generation} layer, coarse-grained metrics like preference rate capture surface alignment~\cite{wang2024understanding}, while fine-grained interaction quality dimensions remain under-specified, see Table~\ref{tab:bench_survey}. \prefix~addresses these gaps by unifying evaluation across all three layers comprehensively.

\subsection{User Experience Evaluation Approaches for Language Agents}

In evaluating user experience, a fundamental methodological divide further complicates assessment \cite{zhao-etal-2025-sphere}. On one hand, traditional HCI evaluation offers qualitative depth through Likert scales and interviews but lacks the reproducibility and scalability required for large-scale comparison \cite{he2025plan,wang2024understanding}. On the other hand, the NLP community achieves scaling at the cost of rigor; user-centric metrics like satisfaction remain fragmented \cite{diebel2025ai}, single faceted \cite{wu2025collabllm}, and are often treated as secondary analyses rather than formalized core objectives. Most importantly, interaction trajectories are usually treated as supplementary material, instead of being established as primary evaluation targets that can be systematically measured and aligned with specific user experience goals and interaction preferences \cite{liu2025human,LogsWu}.\prefix~bridges these gaps by formalizing multi-turn trajectories into unified, reproducible metrics, reconciling HCI-level depth with NLP-level scalability and reproducibility.

\section{\prefix}
\begin{figure*}[ht]
    \centering
    \includegraphics[width=0.9\textwidth]{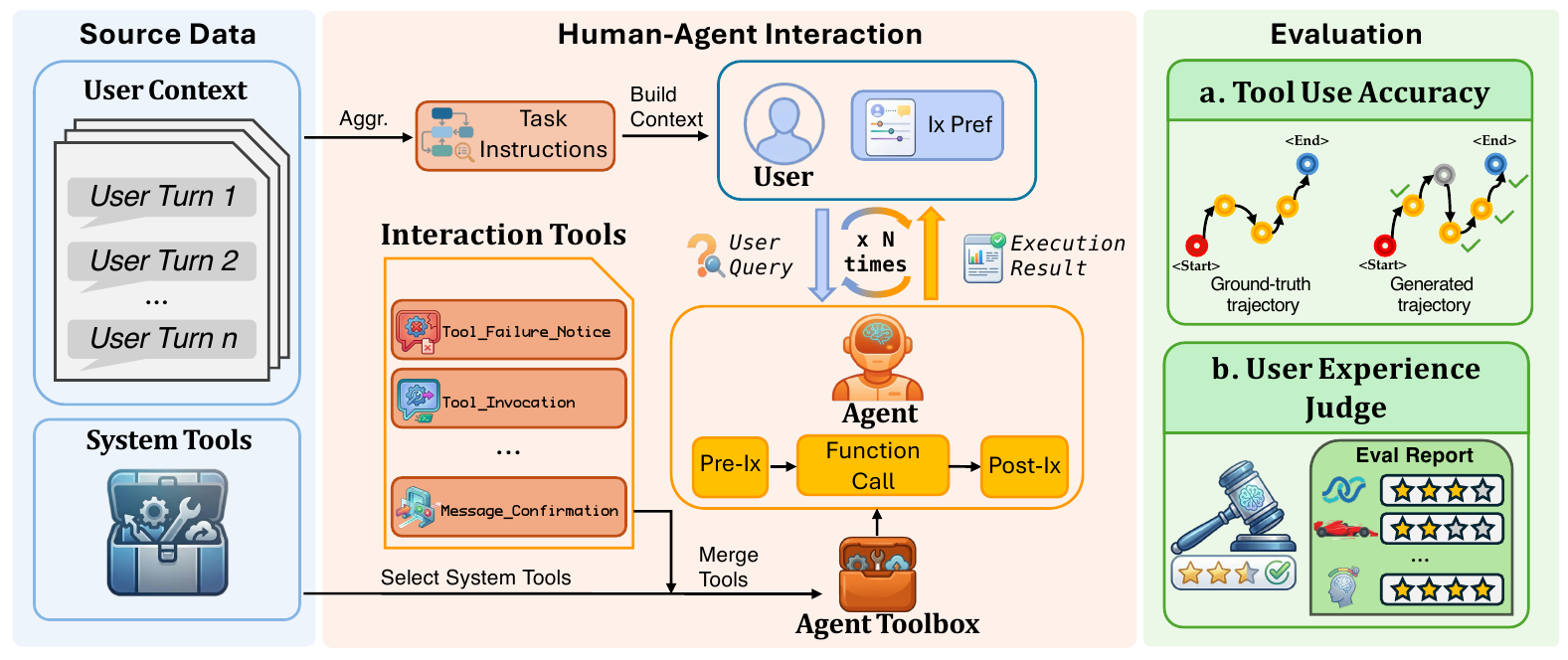}
    \caption{Overview of \prefix. Tasks from BFCL are coarsened into flexible instructions (left), a preference-aware simulator interacts with the agent expressing preferences implicitly (center), and the UX Judge evaluates the resulting trajectory across seven dimensions (right).}
    \label{fig:pipeline}
    \vspace{-0.3cm}
\end{figure*}

To systematically investigate how agents interact with users exhibiting diverse interaction preferences, we introduce \prefix, an interactive evaluation environment that simulates users with varying preference profiles, assessing task accuracy and interaction quality comprehensively. Figure~\ref{fig:pipeline} provides an overview of the construction and evaluation pipeline of \prefix, which consists of four components.

\begin{itemize}[leftmargin=*]

\item[\ding{224}] \textbf{\textit{Task Coarsening.}}
Rewrite BFCL's (Berkeley Function Calling Leaderboard~\cite{patil2025bfcl}) overly-specified prompts into coarser Task Instructions while preserving deterministic tool-use ground truth and enabling flexible multi-turn interactions that elicit and express user preferences.
\vspace{-0.5em}

\item[\ding{224}] \textbf{\textit{Preference-Aware User Simulation.}} 

Define user interaction preferences spanning agent preference adaptation action space  (see Table~\ref{tab:full_hierarchical_taxonomy} in Appendix~\ref{sec:appendix_hierarchical_definitions}); LLM simulators implicitly express assigned preferences through conversational behavior during task without explicit self-disclosure.

\vspace{-0.3em}

\item[\ding{224}] \textbf{\textit{Interaction-as-a-Tool (IaaT).}}
Abstract user-agent interactions as structured tool calls, unifying interaction preferences and task execution within a single measurable framework for quantitative agent alignment benchmarking.

\vspace{-0.3em}

\item[\ding{224}] \textbf{\textit{User Experience (UX) Judge.}}
Move beyond tool-use accuracy by defining a set of core user experience dimensions to assess interaction quality from multiple perspectives. \textit{Interaction Preference Alignment} is assessed by comparing the agent's generated interaction tool invocations against the ground-truth task trajectory associated with the assigned preference label via an LLM judge. For each dimension, the LLM judge outputs: (1) a Likert-scale rating, (2) a justification explaining the score, and (3) specific turn-level evidence demonstrating the observed interaction qualities. Specifically, for each interaction log, we obtain the corresponding user preference setting (e.g., \texttt{each\_confirmation}) and its ground-truth interaction trajectory. 

\end{itemize}

\vspace{-0.3cm}
\subsection{Task Coarsening}
Task prompts in existing benchmarks are generally specified and self-contained, deliberately designed to yield unique solutions or trajectories, thereby offering limited room for differentiated interaction behaviors. To create opportunities for preference-elicited interactions, we first filter tasks that afford meaningful expression of specific preference dimensions (excluding trivial tasks or those irrelevant to targeted preferences). We then employ an LLM to aggregate and rewrite multiple scripted user turns from BFCL into coarser, high-level Task Instructions for the simulator. The rewriting process adheres to two constraints: (1) preserve original task intent and parameter specifications to ensure deterministic tool invocations and maintain ground-truth compatibility for downstream accuracy evaluation; (2) retain the inherent task ordering when explicit sequential dependencies exist (e.g., "first do X, then Y"), while generalizing the phrasing to avoid rigidly prescribing which information must appear within any single dialogue turn. This abstraction enables flexible, multi-turn interactions that expose preference-alignment capabilities without compromising task validity or accuracy.

\vspace{-0.2cm}
\subsection{Preference-Guided User Simulator}
We define 14 user interaction preference attributes across 4 interaction dimensions: \textit{Transparency \& Auditability}, \textit{Interaction Pace \& Flow}, \textit{Strategy \& Initiative}, and \textit{Robustness \& Adaptability} (see Table~\ref{tab:full_hierarchical_taxonomy} in Appendix~\ref{sec:appendix_hierarchical_definitions}). Each attribute comprises 2–3 preference settings, yielding 31 distinct settings that characterize how users prefer to interact with agents. During simulation, each user simulator is assigned one preference setting and is required to adapt its communication style for each turn accordingly. For instance, a user with a "fast interaction pace" preference might convey the entire high-level task instruction in a single concise query. Crucially, the simulator LLM is instructed to \textit{implicitly} express its assigned preference naturally through conversation while strictly avoiding explicit self-disclosure of preference labels (prompt templates available in Appendix~\ref{sec:appendix_prompts}). This ensures agent alignment is tested against authentic, preference-driven behavior rather than artificially telegraphed intentions from users.

\subsection{Tool Design}
\prefix~involves three types of tools. \textit{System tools}, which interact with external systems, are inherited from BFCL \cite{patil2025bfcl}. To support fine-grained control over selected interaction preference, exploration-exploitation trade-offs and error retry strategies, two additional toolboxes were designed following the same architectural principles and functionality as BFCL. Beyond these conventional system tools, we introduce the \textit{Interaction-as-a-Tool} paradigm.

\textbf{Interaction-as-a-Tool (IaaT).} IaaT extends the agent action space by treating user-agent interactions and system tool calls within a unified formalism. This explicit representation addresses the evaluation gap that arises when interaction behaviors are not annotated alongside system actions. By abstracting both interaction tools and system tools as structured tool calls, we enable comprehensive measurement of all agent actions in a comparable and interpretable manner. Trajectory matching is employed to convert previously qualitative, implicit interaction patterns into explicit, token- or action-level prediction tasks. This operationalization anchors interaction preferences as concrete entities, thereby enabling systematic and quantitative benchmarking of agent alignment and closing the gap in assessing agent behaviors involving interaction dimensions.

\textbf{Interaction Tool Types.}
Interaction tools are designed to be invoked before or after system tools, as illustrated in Appendix~\ref{appendix:workflow} (Figure~\ref{fig:Tool_Type}). Our framework defines two categories of interaction tools: \ding{182} \textit{Narrative} tools, which provide explanatory context or transparency without interrupting execution flow, and \ding{183} \textit{Dialogue Control tools}, which explicitly gate task execution pending user confirmation or clarification. A complete list of interaction tools is provided in Appendix~\ref{sec:appendix_interaction_tools} (List~\ref{lst:full_api_schema}).

\begin{table*}[t]
    \caption{Preference adaptation impact on interaction quality metrics (higher is better). Rows are evaluation metrics, columns are models (each split by No\_P and P). \textbf{Gain} row shows the relative improvement ratio.
    % in Avg: $(P/No\_P - 1)\times 100\%$.
    }
    \vspace{-0.3cm}
    \small
    \renewcommand\arraystretch{1.08}
    \setlength\tabcolsep{6pt}
    \definecolor{Gray}{gray}{0.95}
    \definecolor{LightBlue}{rgb}{0.90,0.95,1}
    \definecolor{BlueBright}{rgb}{0.78,0.88,1}
    \definecolor{LightOrange}{rgb}{1,0.87,0.57} % 深一些的橙色
    % Column types for coloring
    % "b" = No_P Avg: Light blue, "B" = P Avg: BRIGHTER blue, "g" = P=Gray, "o" = Gain: orange
    \newcolumntype{b}{>{\columncolor{LightBlue}}c}
    \newcolumntype{B}{>{\columncolor{BlueBright}}c}
    \newcolumntype{g}{>{\columncolor{Gray}}c}
    \newcolumntype{o}{>{\columncolor{LightOrange}}c}
    \begin{center}
    \begin{tabular}{l|c g|c g|c g|c g}
        \hline
        \hline
        \multirow{2}{*}{\textbf{UX Dimension}} &
        \multicolumn{2}{c|}{\textbf{Gemini 3 Flash}} &
        \multicolumn{2}{c|}{\textbf{Claude Opus 4.5}} &
        \multicolumn{2}{c|}{\textbf{Claude Sonnet 4.5}} &
        \multicolumn{2}{c}{\textbf{Kimi K2}} \\
        \cline{2-9}
        & No\_P & P & No\_P & P & No\_P & P & No\_P & P \\
        \hline
        Initiative Timing           & 3.745 & \cellcolor{Gray}4.319 & 3.654 & \cellcolor{Gray}3.871 & 3.298 & \cellcolor{Gray}3.752 & 3.737 & \cellcolor{Gray}3.916 \\
        Interaction Coherence       & 3.684 & \cellcolor{Gray}3.947 & 3.381 & \cellcolor{Gray}3.407 & 2.994 & \cellcolor{Gray}3.229 & 3.643 & \cellcolor{Gray}3.785 \\
        Intent Alignment Drift      & 4.674 & \cellcolor{Gray}4.808 & 4.264 & \cellcolor{Gray}4.317 & 3.989 & \cellcolor{Gray}4.284 & 4.433 & \cellcolor{Gray}4.514 \\
        Commitment Consistency      & 4.593 & \cellcolor{Gray}4.746 & 4.137 & \cellcolor{Gray}4.247 & 3.701 & \cellcolor{Gray}4.044 & 4.216 & \cellcolor{Gray}4.364 \\
        Interaction Efficiency      & 2.956 & \cellcolor{Gray}3.394 & 2.805 & \cellcolor{Gray}2.838 & 2.422 & \cellcolor{Gray}2.663 & 2.945 & \cellcolor{Gray}3.122 \\
        User Cognitive Load Trajectory & 3.283 & \cellcolor{Gray}3.971 & 3.269 & \cellcolor{Gray}3.462 & 2.807 & \cellcolor{Gray}3.249 & 3.304 & \cellcolor{Gray}3.393 \\
        Overall User Experience     & 3.343 & \cellcolor{Gray}4.145 & 3.470 & \cellcolor{Gray}3.777 & 3.079 & \cellcolor{Gray}3.601 & 3.416 & \cellcolor{Gray}3.519 \\
        \textbf{Avg}               
            & \cellcolor{LightBlue}3.754 & \cellcolor{BlueBright}4.190
            & \cellcolor{LightBlue}3.569 & \cellcolor{BlueBright}3.703
            & \cellcolor{LightBlue}3.184 & \cellcolor{BlueBright}3.546
            & \cellcolor{LightBlue}3.671 & \cellcolor{BlueBright}3.802 \\
        \multicolumn{1}{l|}{\textbf{Gain (\%)}}    
            & \multicolumn{2}{o|}{+\;11.6\%}
            & \multicolumn{2}{o|}{+\;3.8\%}
            & \multicolumn{2}{o|}{+\;11.4\%}
            & \multicolumn{2}{o}{+\;3.6\%} \\
        \hline
        \hline
    \end{tabular}
    \end{center}
    \vspace{-0.5cm}
    \label{tab:personalization_main_2}
\end{table*}

\subsection{Evaluation}
\subsubsection{Tool Use Accuracy}
The tool-use accuracy score (i.e. Subset-Matched Response-based Evaluation) is retained from BFCL v3 multiturn setting to evaluate agents' performance on traditional task correctness. The \textit{tool-use accuracy} is defined as:
\[
\frac{|\text{Generated Trajectory} \cap \text{GT Trajectory}|}{|\text{GT Trajectory}|}  (\%)
\]
where the numerator represents the number of correctly predicted tool calls (including function names and parameters) that match the ground truth trajectory, and the denominator is the total number of tool calls in the reference trajectory.

\subsubsection{User Experience Judge (UX Judge)}
We employ a composite LLM-as-judge framework with multiple judges to assess interaction experience across seven complementary dimensions (UX dimension) and an extra dimension for interaction preference alignment. Each judge provides Likert-scale ratings from 1 to 5 to quantify distinct aspects of interaction quality. Ground-truth interaction trajectories, specifying the expected sequence of interaction tool calls for each preference setting, are provided in Appendix~\ref{sec:appendix_interaction_tools}.

\textbf{Initiative timing.} Initiative timing denotes when and how an interactive system takes initiative during an interaction \cite{Horvitz_init_1}. Poorly timed initiative can markedly degrade user performance \cite{McFarlane_init_2}. In human-agent interaction, proper timing means the LLM agent proposes actions or interruptions at opportune moments neither prematurely (causing disruption or annoyance) nor belatedly (missing the moment of need) \cite{abbas2022understanding,peng2024human}. 

\textbf{Interaction Coherence.} Interaction coherence captures the logical consistency and connectedness of an ongoing exchange. A coherent human-agent interaction keeps the agent’s contributions contextually sensible, especially over long contexts typical for LLM agents. The agent recalls prior events introduced by the user, avoids off-topic responses \cite{SINGH20228852}, and refrains from abrupt topic shifts. High coherence yields smooth, easy-to-follow dialogue, whereas incoherence (contradictions, entangled messages) confuses users \cite{maharjan2022s_coh_1}.

\textbf{Intent Alignment Drift.} Intent alignment means the system correctly infers and remains aligned with the user’s goals, preferences, and intents over time. Drift arises when lengthy context induces attention decay, causing the agent to lose track of foundational constraints \cite{kim2024understanding}, or when the agent fixates on a subtask and neglects the overarching objective or parallel subgoals. Agents that minimize intent alignment drift are perceived as more cooperative, useful, and trustworthy \cite{attig2025understanding}.

\textbf{Commitment Consistency.} Commitment consistency requires the system to honor implied or explicit commitments and behave in line with user expectations\cite{cila2022designing}. Once a plan, choice, or rule (“commitment”) is established, the system should not contradict or deviate without justification. In human-agent interaction, consistent agents exhibit clear state tracking, verification loops before presenting outputs, and proactive confirmation when deviations are unavoidable, handling errors with backup, risk-mitigated plans. Consistency builds predictability and trust—sharpening users’ mental models of agent capabilities—whereas breaking commitments or behaving unreliably undermines confidence \cite{daronnat2021inferring}.

\textbf{Interaction Efficiency.} Interaction efficiency, a core usability pillar in HCI (e.g., ISO 9241) \cite{draper1993notion}, measures how quickly and effortlessly users achieve goals with the system \cite{ding-etal-2023-harnessing}. An efficient agent minimizes unnecessary steps, delays, and cognitive effort. High efficiency delivers value with speed and minimal friction, whereas low efficiency correlates negatively with perceived competence or preference for the system \cite{abbas2022understanding}. Efficiency strongly influences user satisfaction, particularly for goal-directed task completion.

\textbf{Cognitive Load.} Grounded in cognitive load theory \cite{hollender2010integrating}, this dimension addresses the limits of working memory and how system demands can exceed them. Cognitive load captures the mental effort needed to perform a task relative to working-memory capacity. High load impairs performance and learning, whereas optimal load enables efficient processing \cite{daronnat2021inferring}. In human-agent interaction, agents with predictable behavior reduce reported cognitive load, and increased predictability improves trust and overall task performance.

\textbf{Interaction Preference Alignment.} Interaction Preference Alignment assesses how effectively an agent’s autonomy and initiative, information density and modality, decision-making logic, and communicative style match, adapt to, and remain consistent with a user’s stated or implicit preferences, thereby maintaining user comfort \cite{goyal2024designing}. This alignment should persist across the full collaborative task space, ensuring that the “how” of interaction is as satisfactory as the “what” of the task outcome. Strong alignment boosts trust, satisfaction, fluency, and reduces clarifications or rework. Poor alignment causes frustration, cognitive load, detours, or task abandonment despite competent execution.

\textbf{Overall User Experience.} Overall UX provides an aggregate assessment of the user's interaction experience with the agent, encompassing reuse intention, perceived trust, interaction smoothness, and perceived reliability (i.e., whether the interaction is satisfactory without being intrusive or annoying). This dimension serves as a holistic summary of the seven individual UX dimensions defined above.

\subsection{Configuration}

The benchmark composition, including task distribution and tool inventory, is detailed in Table~\ref{tab:bench_stats} in Appendix~\ref{appendix:stats}. Our preference attributes are designed to be atomic and configurable, enabling flexible random combinations for diverse user profiles. It provides additional possibilities for exploring compositional preferences by combining multiple attributes to simulate more naturalistic, complex user behaviors and increase evaluation challenge. The design of interaction tools follows a \textbf{minimalist} approach: when similar interaction preferences can be expressed through different parameter configurations of a single tool, we avoid introducing additional tools. This modular design ensures extensibility—new interaction tools can be seamlessly integrated as novel interaction patterns or optimization opportunities emerge in agent action spaces.
\section{Experiment}
\subsection{Preference Adaptation Setup}
We compare two conditions: \textbf{Adaptation} (denoted as P), where models receive instructions to infer preferences from a sample interaction history containing implicit user feedback signals, and \textbf{Baseline} (denoted as No\_P), where models receive generic instructions without history. Both conditions use identical tools and execution loops.

\subsection{Experimental Configuration}
Four models were selected for their function-calling capabilities: 
\texttt{claude-opus-4-5}~\cite{anthropic2025opus45}, \texttt{claude-sonnet-4-5}~\cite{anthropic2025sonnet45},
\texttt{gemini-3-flash-preview}~\cite{google2025gemini3}, and \texttt{kimi-k2-0905-preview}~\cite{team2025kimi}. Following BFCL's default configuration, all models use temperature $= 0.1$. Each adaptation condition runs on 283 samples spanning 31 preference settings (average: 9 samples per setting). Tasks are sourced from \texttt{BFCL\_v3\_multi\_turn\_long\_context} and \texttt{BFCL\_v3\_multi\_turn\_miss\_param}, then filtered and coarsened to yield diverse preference-differentiated test cases.

For each task, models receive a system prompt, the user's current turn query, and a list of available tools. Models must complete tasks within 180 message exchanges (including tool invocations); exceeding this limit results in termination, though partial progress is recorded and evaluated. Tool Use Accuracy is computed from the sequence of generated tool invocations, while UX evaluation requires the complete conversation history, which is retained across all experimental runs.

\subsection{Simulator and Judge Configuration}
User simulation employs \texttt{gpt-4.1}~\cite{achiam2023gpt} with temperature $= 0$. For UX evaluation, each of the four evaluated models serves as an independent judge with temperature $= 0$, assessing interaction quality without cross-contamination.

\section{Results}
We evaluate \prefix~along two axes: (1) whether incorporating interaction tools preserves task-level accuracy, and (2) whether preference-aware adaptation improves user experience metrics.

\subsection{Baseline Validation and General Trends}

\begin{figure}[t]
    \centering
    \includegraphics[width=0.85\linewidth]{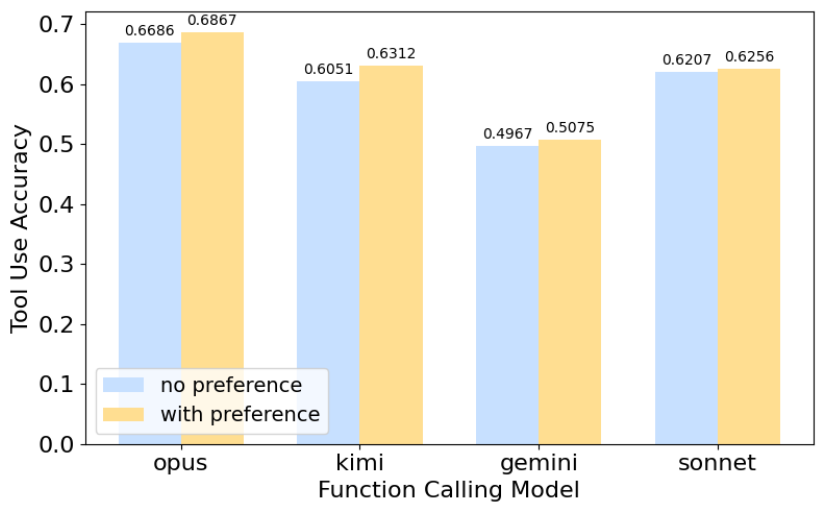}
    \caption{Tool-use accuracy comparison between Baseline (No\_P) and Adaptation (P) conditions across four models. Baseline performance is comparable to BFCL leaderboard scores, confirming that interaction tools do not interfere with system tool execution.}
    \label{fig:tu_acc}
\end{figure}

\textbf{Adding Interaction Tool does not interfere with system tool usage.}  We first evaluate the Baseline condition to verify that interaction tools do not degrade core functional capabilities. As shown in Figure~\ref{fig:tu_acc}, baseline performance is comparable to scores on BFCL leaderboard long-context settings, confirming that adding interaction tools does not compromise core system tool execution.

The same figure also shows that, across all four models, the Adaptation condition consistently outperforms the Baseline, demonstrating that preference adaptation enhances tool-use accuracy without compromising core execution.

\textbf{Universal Tool Accuracy Gains.} Adaptation causes broadly consistent accuracy gains. In preference breakdown, the largest gains occur in \texttt{info\_collect\_upfront} ($\text{mean\_lift} = 0.083$). Only 3 of 31 preference settings see more than 3 models degraded, with degradation typically under 2\%. The only outlier is \texttt{tool\_invocation\_multiple}, where multi-invocation interaction preferences might have increased difficulty and thus interfered with tool execution logic.

Having established that interaction tools preserve task accuracy, we now examine whether preference adaptation improves overall user experience.

\subsection{User Experience Judge}
\begin{table}[t]
    \caption{Interaction Preference Alignment scores with and without personalization (higher is better). 
    % Each row is a model, columns are No\_P/P (gray: P), with Avg and Gain (\%).
    }
    \vspace{-0.3cm}
    \small
    \renewcommand\arraystretch{1.10}
    \setlength\tabcolsep{7pt}
    \definecolor{Gray}{gray}{0.95}
    \definecolor{LightBlue}{rgb}{0.90,0.95,1}
    \definecolor{BlueBright}{rgb}{0.78,0.88,1}
    \definecolor{LightOrange}{rgb}{1,0.87,0.57}
    % Column coloring
    \newcolumntype{g}{>{\columncolor{Gray}}c}
    \newcolumntype{o}{>{\columncolor{LightOrange}}c}
    \begin{center}
    \begin{tabular}{l|c g|c}
        \hline
        \hline
        \textbf{Model} 
        & \textbf{No\_P} 
        & \textbf{P}
        & \textbf{Gain (\%)} \\
        \hline
        Gemini 3 Flash 
            & 3.142 & \cellcolor{Gray}4.152
            & +32.2 \\
        Claude Sonnet 4.5    
            & 3.210 & \cellcolor{Gray}3.930
            & +22.4 \\
        Claude Opus 4.5      
            & 3.429 & \cellcolor{Gray}3.983
            & +16.2 \\
        Kimi K2         
            & 3.324 & \cellcolor{Gray}3.461
            & +4.1 \\
        \hline
        \textbf{Avg}
            & \cellcolor{LightBlue}3.276
            & \cellcolor{BlueBright}3.882
            & \cellcolor{LightOrange}+18.5 \\
        \hline
        \hline
    \end{tabular}
    \end{center}
    \label{tab:interaction_alignment}
    \vspace{-0.5cm}
\end{table}

Table~\ref{tab:personalization_main_2} demonstrates consistent improvement across all seven UX dimensions, with an average gain of 7.6\%. Among dimensions, \textit{initiative timing} and \textit{overall user experience} exhibit the largest absolute gains, while \textit{interaction efficiency}—despite lower baseline scores—shows meaningful improvement. Across models, Gemini 3 Flash (+11.6\%) and Claude Sonnet 4.5 (+11.4\%) benefit most from adaptation, whereas Claude Opus 4.5 and Kimi K2 show more modest gains (+3.6--3.8\%), suggesting that stronger baseline models have limited room for further improvement.

Most notably, \textit{interaction preference alignment} (Table~\ref{tab:interaction_alignment}) shows a substantial 18.5\% average improvement, with Gemini 3 Flash achieving +32.2\%. This suggests that providing interaction history and preference-aware queries enables models to more effectively execute behavioral constraints.

\begin{figure}[t]
    \centering
    \includegraphics[width=0.7\linewidth]{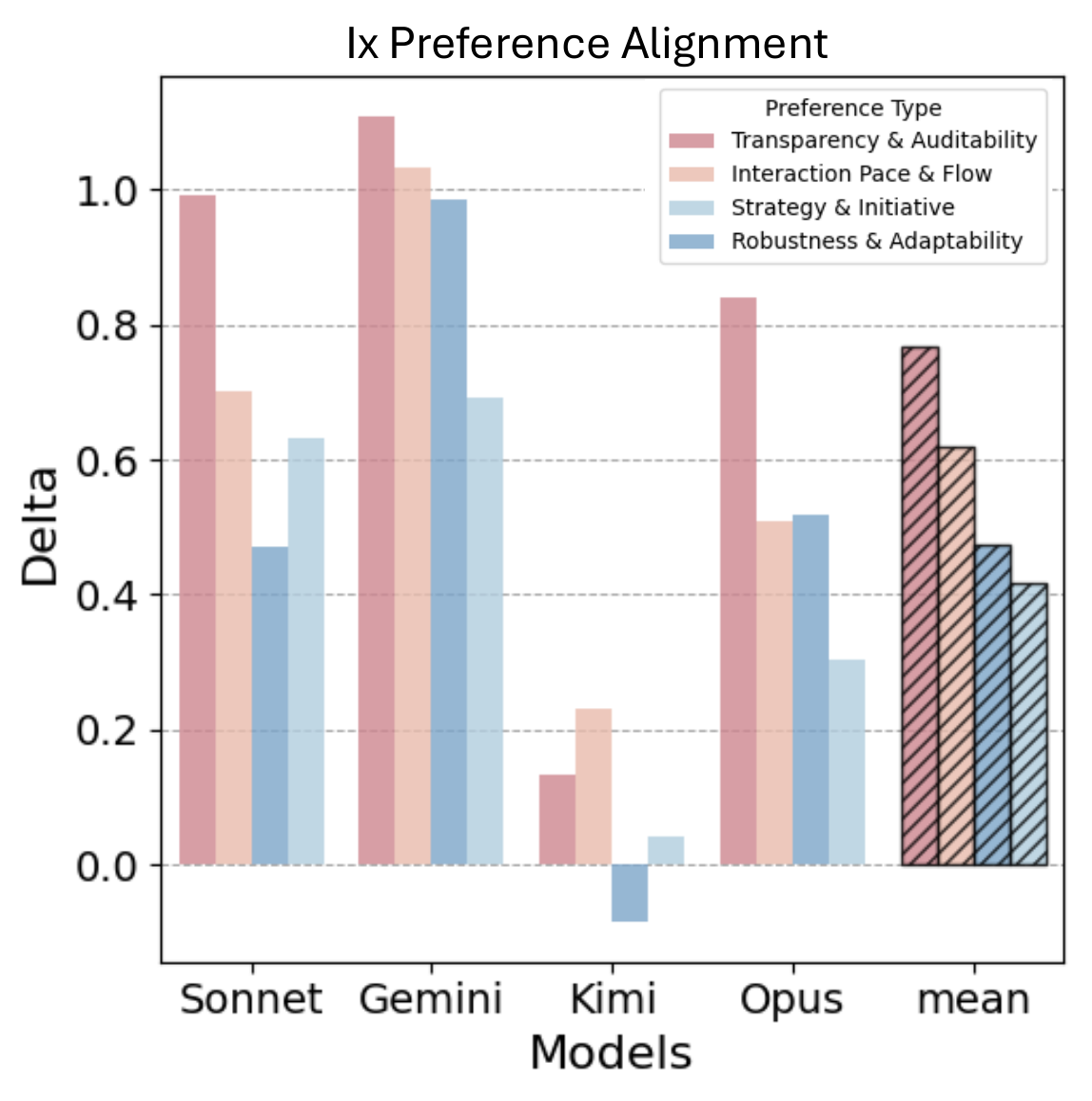}
    \vspace{-0.3cm}
    \caption{Performance gains in interaction preference alignment across four categories, measured as the delta between adaptation and baseline conditions. Results are aggregated by category and normalized by the number of preference settings per group.}
    \label{judge_ix_preference}
\end{figure}

Breaking down by preference category (Figure~\ref{judge_ix_preference}), \textit{transparency \& auditability} consistently achieves the largest alignment gains across most models (Sonnet, Gemini, Opus), suggesting that current agents are more capable of aligning with transparency-related preferences. While \textit{interaction pace \& flow} also shows substantial improvements, \textit{strategy \& initiative} and \textit{robustness \& adaptability} exhibit relatively lower deltas, indicating that preferences requiring holistic adjustments to global interaction patterns prove harder to align than localized behavioral changes.

Figure~\ref{judge_more} shows how preference categories differentially impact user experience metrics. \textit{Robustness \& adaptability} drives \textit{interaction efficiency} gains with the most dominant result, while \textit{transparency \& auditability} dominates \textit{cognitive load} and \textit{initiative timing} improvements.

These findings enable modular agent optimization: rather than treating alignment monolithically, practitioners can prioritize specific preference categories for targeted UX improvements. For example, given that \textit{robustness \& adaptability} is the primary contributor to perceived interaction efficiency, reinforcing models' capability on tool change preference alignment could effectively enhance overall perceived efficiency. Consequently, the alignment process can be tailored to an agent's specific functional goals, leveraging these dominant preference categories as key intervention points.

\section{Analysis}
% =============== 图片导入区===============
% \begin{table}[ht]
%     \centering
%     \renewcommand{\arraystretch}{1.08}
%     \small
%     \setlength{\tabcolsep}{5pt}
%     \begin{tabular}{lcc}
%     \hline
%     \textbf{Dimension} & \textbf{ICC$_{2,1}$} & \textbf{ICC$_{2,k}$} \\
%     \hline
%     \texttt{initiative timing} & 0.495 & 0.797 \\
%     \texttt{interaction coherence} & 0.569332 & 0.840964 \\
%     \texttt{intent alignment drift} & 0.501313 & 0.800839 \\
%     \texttt{commitment consistency} & 0.521520 & 0.813426 \\
%     \texttt{interaction preference alignment} & 0.491620 & 0.794583 \\
%     \texttt{interaction efficiency} & 0.578110 & 0.845706 \\
%     \texttt{user cognitive load trajectory} & 0.506546 & 0.804157 \\
%     \texttt{overall user experience} & 0.507858 & 0.804982 \\
%     \hline
%     \end{tabular}
%     \caption{Inter-rater reliability (ICC) for each evaluation dimension.}
%     \label{tab:icc_scores}
% \end{table}

\vspace{-0.4cm}

\begin{table}[ht]
    \caption{Inter-rater reliability (ICC) for each dimension.}
    \vspace{-0.2cm}
    \centering
    \small
    \setlength{\tabcolsep}{5pt}
    \begin{tabular}{lcc}
    \hline
    \textbf{Dimension} & \textbf{ICC$_{2,1}$} & \textbf{ICC$_{2,k}$} \\
    \hline
    \texttt{initiative timing} & 0.495 & 0.797 \\
    \texttt{interaction coherence} & 0.569 & 0.841 \\
    \texttt{intent alignment drift} & 0.501 & 0.801 \\
    \texttt{commitment consistency} & 0.522 & 0.813 \\
    \texttt{interaction preference alignment} & 0.492 & 0.795 \\
    \texttt{interaction efficiency} & 0.578 & 0.846 \\
    \texttt{user cognitive load trajectory} & 0.507 & 0.804 \\
    \texttt{overall user experience} & 0.508 & 0.805 \\
    \hline
    \end{tabular}
    \vspace{-0.4cm}
    \label{tab:icc_scores}
\end{table}

\begin{table}[ht]
    \caption{Mean standard deviation and coefficient of variation (CV) across models.}
    \vspace{-0.2cm}
    \centering
    \small
    \renewcommand{\arraystretch}{1.15}
    \begin{tabular}{lcc}
    \hline
    \textbf{Model} & \textbf{Mean Std} & \textbf{Mean CV} \\
    \hline
    claude sonnet 4.5 & 0.03429 & 0.88\% \\
    gemini 3 flash & 0.04242 & 1.64\% \\
    kimi k2 & 0.09042 & 2.23\% \\
    claude opus 4.5 & 0.00000 & 0.00\% \\
    \hline
    \end{tabular}
    \label{tab:mean_std_cv}
    \end{table}

\begin{figure}[t]
    \centering
    \includegraphics[width=0.85\linewidth]{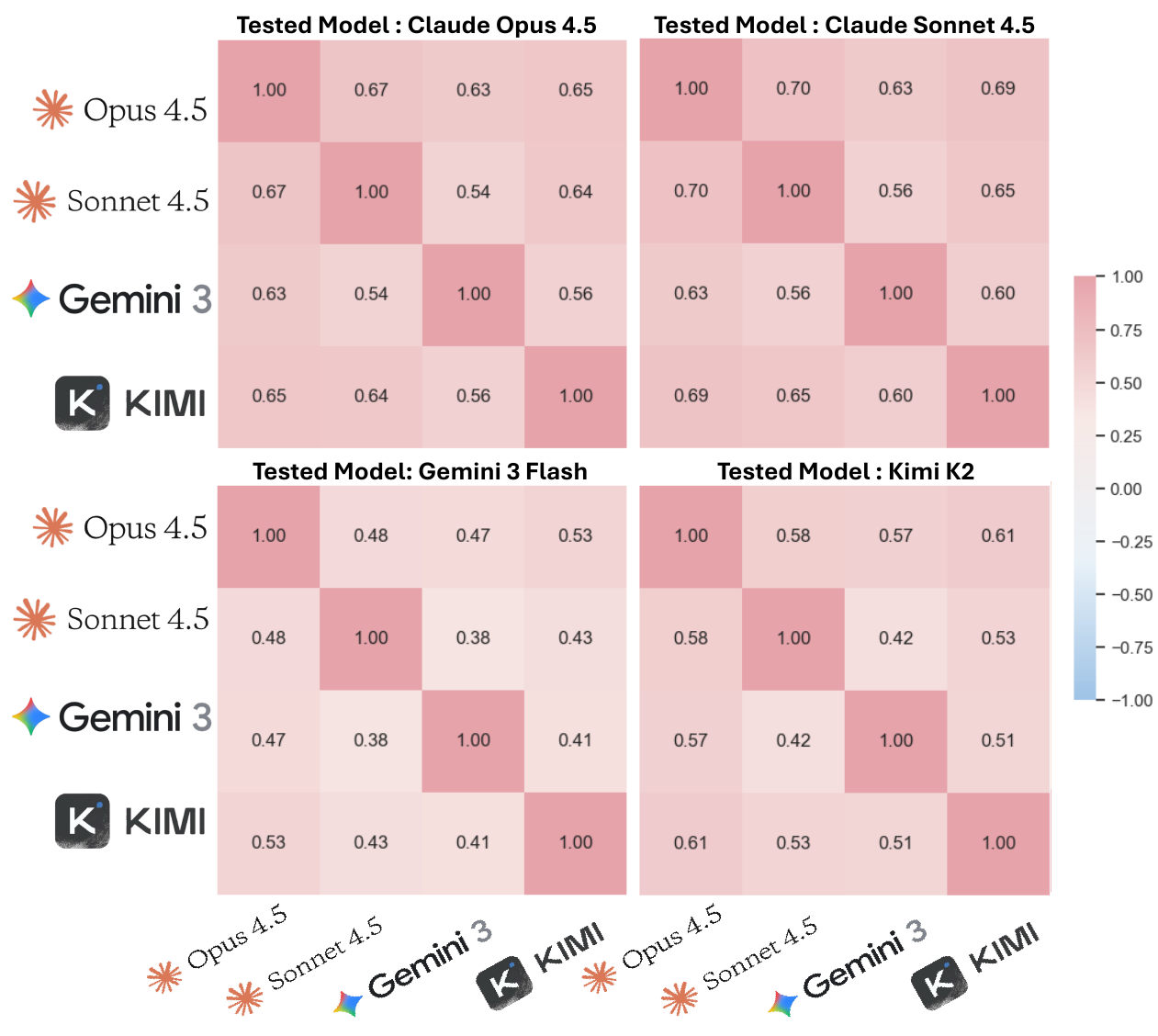}
    \caption{Inter-judge correlation heatmap for UX scores. Each cell shows the Pearson correlation between two LLM judges. Moderate correlations (0.4--0.7) indicate a shared underlying construct while preserving diverse perspectives.}
    \label{fig:heatmap}
    \vspace{-0.5cm}
\end{figure}
% =============== 图片导入区===============

\textbf{Multi-Judge for Robust UX Evaluation.}
A key characteristic of \prefix~is multiple LLMs as UX judges. This multi-judge design addresses two critical measurement challenges: robustness and bias mitigation. Aggregating judgments reduces stochastic errors from prompt sensitivity and hallucinations, while leveraging diverse model biases produces more balanced overall assessments.

Table~\ref{tab:icc_scores} shows inter-rater reliability via ICC scores~\cite{ICC}. ICC(2,1) (single judge reliability) remains below 0.57 across all seven dimensions, indicating instability. However, ICC(2,k) (aggregate reliability) consistently exceeds 0.79. This demonstrates that while individual LLM judges are unreliable in isolation, their collective judgment achieves strong reliability, justifying the multi-judge framework's necessity.
% demonstrating that collective judgment achieves strong reliability despite individual unreliability. 

Figure~\ref{fig:heatmap} shows pairwise correlations from 0.4 to 0.7, indicating diverse perspectives. Averaging this heterogeneity yields consensus with substantially higher reliability than individual judgments, validating multi-judge design.

\textbf{Internal Consistency of UX Dimensions.}
Cronbach's alpha across the seven UX dimensions was 0.943 (95\% CI: [0.939, 0.946]). This excellent reliability confirms that our dimensions measure distinct facets yet converge on a unified UX construct with strong coherence.

\textbf{Reproducibility.}
We evaluated reproducibility through 20 independent runs on 5 representative samples. Table~\ref{tab:mean_std_cv} shows mean CV (Coefficient of Variation) values for each model, demonstrating strong within-judge stability. Combined with multi-judge averaging, these results confirm the reproducibility of our UX Judge design.

\textbf{Human Validation.}
Lastly, to validate automated scores against human judgment, 5 volunteers independently rated 20 sampled interactions (80 judge-model pairs total) using identical criteria without seeing LLM scores. Spearman correlations between averaged human ratings and LLM scores range from 0.52–0.78 across dimensions, with strongest agreement on Cognitive Load ($\rho = 0.78$), and Interaction Preference Alignment ($\rho = 0.75$) . Humans and LLMs frequently cited identical turns as evidence, indicating convergent interpretation of interaction quality. This validates LLM judges' fidelity for large-scale UX evaluation.
\section{Conclusion}
We introduce \prefix, a configurable environment for evaluating LLM-based agent on aligning human interaction preferences. Proposing the Interaction-as-a-Tool (IaaT) paradigm, we formalize interaction alignment as a core metric of equal importance to task accuracy, supported by a structured taxonomy of agent trajectories. Furthermore, we address the challenge of reproducibility by establishing a composite LLM-as-a-Judge mechanism. Spanning seven distinct yet coherent UX dimensions, this mechanism transforms subjective interaction experiences into unified metrics, ensuring traceable, reproducible, reliable, and scalable assessment across dynamic multi-turn interaction scenarios.

\section*{Limitations}
First, regarding simulation fidelity, while our simulated users are effective, they may not fully capture the spectrum of behavioral diversity and nuanced interaction preferences found in real human populations. Second, regarding task complexity, our current study focuses on distinct preference types; future work should explore the compound effects and increased difficulty of satisfying multiple, potentially conflicting preferences simultaneously. Third, to enrich evaluation perspectives, we plan to employ simulators to generate "self-reported" satisfaction scores alongside interaction data, allowing for comparative analysis between these intrinsic simulated reports and external evaluations (such as our LLM Judge). Finally, we observed a trade-off whereby adapting to specific preferences can occasionally compromise tool-use accuracy due to increased control demands. To address this, we aim to leverage interaction preference alignment as a reinforcement learning signal, training agents to balance personalization with functional precision in user-experience-sensitive domains.
\section*{Ethics Statement}
This work evaluates LLM-based agents through personalized, multi-turn interactions, focusing on interaction quality and preference adaptation. While primarily methodological, it raises ethical considerations regarding broader impacts, fairness, and responsible use.

This research benefits developers of interactive agents by enabling systematic, user-centric evaluation. However, interaction optimization could be misused to manipulate user compliance or steer behavior, particularly in high-stakes domains like decision support. Personalization should prioritize user alignment over engagement maximization.

All data are generated through controlled simulations; no human data are collected, ensuring no privacy concerns or consent requirements. However, underlying models may encode social or linguistic biases affecting agent behavior and LLM judgments. Preference dimensions are not value neutral and may not generalize across cultural contexts. Future work should expand preference coverage and assess potential disadvantages to specific user groups.

Simulated evaluation cannot capture nuanced human reactions like long-term trust or frustration. This framework complements, but does not replace, human-centered studies in deployment settings. We document the framework's scope and limitations, emphasizing transparent reporting and conservative deployment with user control mechanisms.

\bibliography{custom}
\newpage

\appendix
\newpage
\clearpage
\onecolumn
% \section{User Preferences}
% \input{Appendix_Tabs/Pref_List Complete.tex}

%  initiative_timing， interaction_coherence，intent_alignment_drift，commitment_consistency，interaction_efficiency，user_cognitive_load_trajectory，overall_user_experience
% 和他们具体的定义跟likert

\section{Interaction Preference Taxonomy}

\begin{figure*}[h!]
    \centering
    \includegraphics[width=\textwidth]{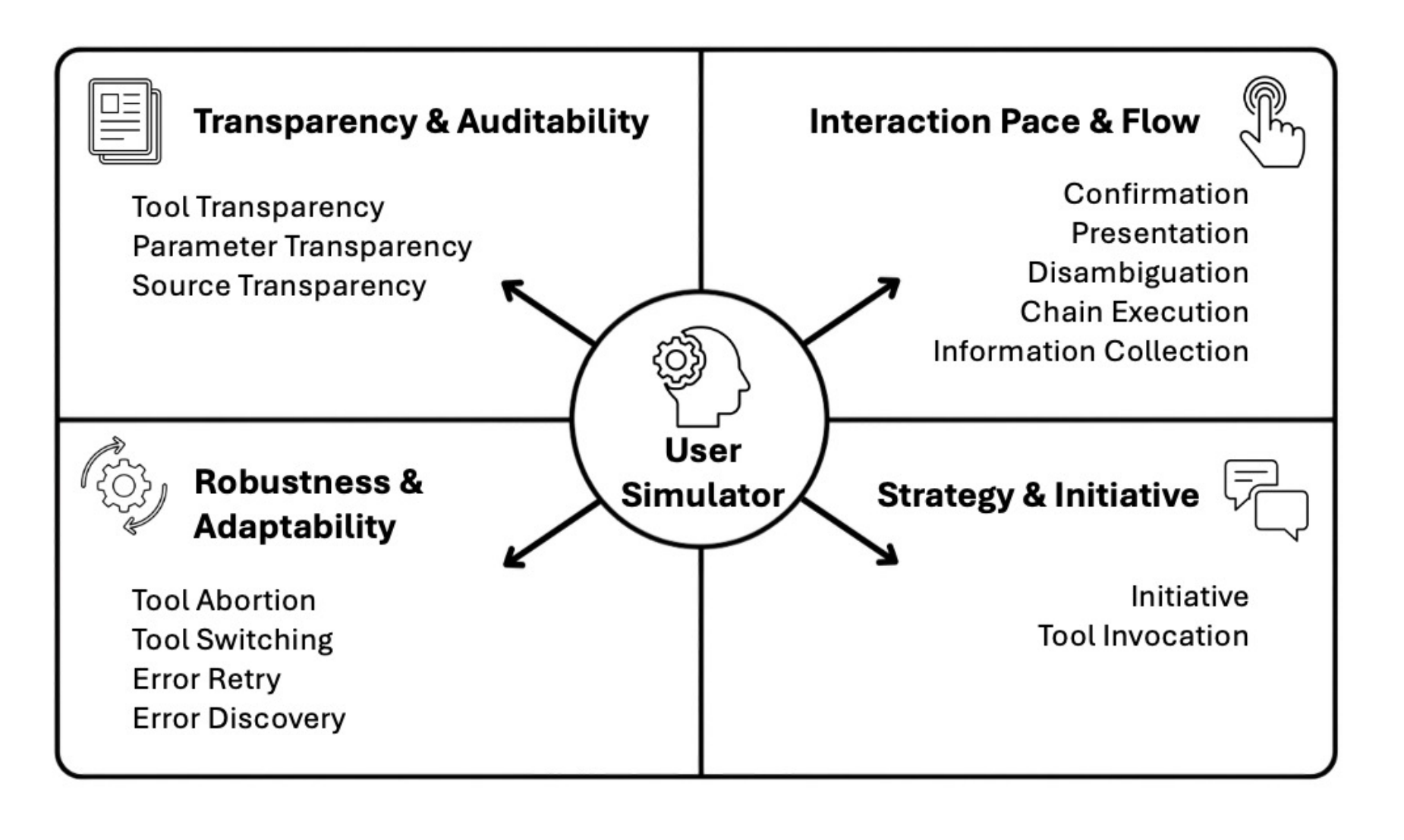}
    \caption{Hierarchical structure of interaction preference dimensions, attributes, and settings.}
    \label{fig:preference_dimensions}
\end{figure*}

\label{sec:appendix_hierarchical_definitions}

\begin{table*}[h!]
\centering
\scriptsize % 字号设为脚本大小，以容纳大量文字
\renewcommand{\arraystretch}{1.2} % 调整行间距
\begin{tabularx}{\textwidth}{@{} p{2cm} p{2cm} p{1.6cm} X @{}}
\toprule
\textbf{Dimension} & \textbf{Attribute} & \textbf{Setting} & \textbf{Definition} \\
\midrule

% ==========================================
% 1. Transparency & Auditability (8 rows)
% ==========================================
\multirow{8}{2cm}{\textbf{Transparency \& Auditability}} 
 & \multirow{3}{2cm}{Tool Transparency} & High & Wants explicit tool choice and reasoning before execution (no gate unless combined with confirmation). \\
 & & Medium & Prefers a brief mention of tool choice without gating; wants context but not friction. \\
 & & Low & Prefers silent tool choice/execution; views tooling as internal details. \\
 \cmidrule(l){2-4}
 & \multirow{3}{2cm}{Parameter Transparency} & High & Wants parameter names/values and rationale shown before execution. \\
 & & Medium & Wants light visibility into key parameters but no stepwise approval. \\
 & & Low & Prefers autonomous parameter selection with no exposure of values or rationale. \\
 \cmidrule(l){2-4}
 & \multirow{2}{2cm}{Source Transparency} & High & Wants sources cited; rejects opaque answers. \\
 & & Low & Prefers answers without source exposition unless requested. \\
\midrule

% ==========================================
% 2. Interaction Pace & Flow (11 rows)
% ==========================================
\multirow{11}{2cm}{\textbf{Interaction Pace \& Flow}} 
 & \multirow{3}{2cm}{Confirmation} & Each & Requires confirmation for every individual action; prioritizes safety and situational awareness. \\
 & & Silent & Wants automatic execution without asking; prioritizes speed and minimal friction. \\
 & & Batch & Prefers one confirmation for a related group of actions instead of per-step gating. \\
 \cmidrule(l){2-4}
 & \multirow{2}{2cm}{Presentation} & Compact & Prefers concise, linear output; low tolerance for verbosity. \\
 & & Layered & Prefers layered/expandable output: summary first, details on demand. \\
 \cmidrule(l){2-4}
 & \multirow{2}{2cm}{Info Collection} & Upfront & Prefers all required info requested in one bundle before proceeding. \\
 & & Gradual & Wants required info gathered stepwise, not all at once. \\
 \cmidrule(l){2-4}
 & \multirow{2}{2cm}{Disambiguation} & Upfront & Prefers all ambiguity resolved in one shot to avoid repeated interruptions. \\
 & & Gradual & Prefers clarifications to arrive incrementally rather than a large upfront list. \\
 \cmidrule(l){2-4}
 & \multirow{2}{2cm}{Chain Execution} & Parallel & Prefers parallel execution for speed when tasks are independent. \\
 & & Sequential & Prefers stepwise execution with intermediate visibility. \\
\midrule

% ==========================================
% 3. Strategy & Initiative (4 rows)
% ==========================================
\multirow{4}{2cm}{\textbf{Strategy \& Initiative}} 
 & \multirow{2}{2cm}{Initiative} & Proactive & Wants the agent to act within scope without waiting for every nudge. \\
 & & Reactive & Wants the agent to wait for explicit go-ahead before acting. \\
 \cmidrule(l){2-4}
 & \multirow{2}{2cm}{Tool Invocation} & Single & Prefers picking the best single tool/option over exploring many. \\
 & & Multiple & When available, prefers running multiple options to compare outcomes. \\
\midrule

% ==========================================
% 4. Robustness & Adaptability (8 rows)
% ==========================================
\multirow{8}{2cm}{\textbf{Robustness \& Adaptability}} 
 & \multirow{2}{2cm}{Tool Abortion} & Stop & On failure, wants the workflow to halt instead of continuing. \\
 & & Continue & On partial failure, wants remaining subtasks to continue. \\
 \cmidrule(l){2-4}
 & \multirow{2}{2cm}{Tool Switching} & High Agency & Wants automatic tool switching on failure without asking. \\
 & & Low Agency & Wants to be informed and approve before switching tools. \\
 \cmidrule(l){2-4}
 & \multirow{2}{2cm}{Error Retry} & Silent & Prefers silent, autonomous retries unless failures persist. \\
 & & Escalation & Wants errors surfaced and confirmation before retrying. \\
 \cmidrule(l){2-4}
 & \multirow{2}{2cm}{Error Discovery} & Brief & Wants minimal failure notice; rejects verbose diagnostics. \\
 & & Detail & Wants reasoning/root cause when errors occur. \\

\bottomrule
\end{tabularx}
\caption{Complete taxonomy of interaction dimensions, attributes, and preference settings.}
\label{tab:full_hierarchical_taxonomy}
\end{table*}

% prompts
\UseRawInputEncoding
% --- Appendix: User Simulator Prompt ---

% 切换为单栏模式以获得更好的阅读体验
\clearpage
\onecolumn

\section{Prompt Templates}
\label{sec:appendix_prompts}

% --- 定义 JSON 样式（美化版） ---
\definecolor{json_key}{RGB}{20,105,176}      % 蓝色（属性名）
\definecolor{json_string}{RGB}{18,110,44}    % 绿色（字符串内容）
\definecolor{json_punct}{RGB}{153,51,51}     % 暗红色（标点）

\lstdefinelanguage{json}{
    basicstyle=\ttfamily\small\color{black},
    breaklines=true,
    showstringspaces=false,
    % 关键：识别引号内的字符串颜色
    stringstyle=\color{json_string},
    morestring=[b]",
    morestring=[d]',
    % 设置标点符号颜色
    literate=
     *{0}{{{\color{black}0}}}{1}
      {1}{{{\color{black}1}}}{1}
      {2}{{{\color{black}2}}}{1}
      {3}{{{\color{black}3}}}{1}
      {4}{{{\color{black}4}}}{1}
      {5}{{{\color{black}5}}}{1}
      {6}{{{\color{black}6}}}{1}
      {7}{{{\color{black}7}}}{1}
      {8}{{{\color{black}8}}}{1}
      {9}{{{\color{black}9}}}{1}
      {:}{{{\color{json_punct}{:}}}}{1}
      {,}{{{\color{json_punct}{,}}}}{1}
      {\{}{{{\color{json_key}{\{}}}}{1}
      {\}}{{{\color{json_key}{\}}}}}{1}
      {[}{{{\color{json_key}{[}}}}{1}
      {]}{{{\color{json_key}{]}}}}{1},
}

% --- 终极美化版 JSON 配色 ---
\definecolor{json_key}{RGB}{20,105,176}      % 蓝色：用于 Key
\definecolor{json_string}{RGB}{18,110,44}    % 绿色：用于 Value 内容
\definecolor{json_punct}{RGB}{153,51,51}     % 暗红色：用于标点和括号

\lstdefinelanguage{json_pretty}{
    basicstyle=\ttfamily\small\color{black},
    breaklines=true,
    showstringspaces=false,
    stringstyle=\color{json_string},
    morestring=[b]",
    % 使用 literate 强制替换 Key 的颜色
    literate=
     {"persona"}{{{\color{json_key}\bfseries "persona"}}}{9}
     {"description"}{{{\color{json_key}\bfseries "description"}}}{13}
     {"sample behaviors"}{{{\color{json_key}\bfseries "sample behaviors"}}}{18}
     {"interaction instruction for simulator"}{{{\color{json_key}\bfseries "interaction instruction for simulator"}}}{40}
     {:}{{{\color{json_punct}{:}}}}{1}
     {,}{{{\color{json_punct}{,}}}}{1}
     {\{}{{{\color{json_punct}{\{}}}}{1}
     {\}}{{{\color{json_punct}{\}}}}}{1}
     {[}{{{\color{json_punct}{[}}}}{1}
     {]}{{{\color{json_punct}{]}}}}{1},
}
\lstset{
    basicstyle=\ttfamily\small,
    breaklines=true,
    breakatwhitespace=true,
    frame=single,
    rulecolor=\color{gray!30},
    backgroundcolor=\color{gray!5},
    % --- 核心修改部分 ---
    framesep=1em,      % [内边距] 控制代码与边框之间的距离（相当于 padding）
    xleftmargin=1em,   % [左外边距] 整个代码块距离左侧页边的距离
    xrightmargin=1em,  % [右外边距] 整个代码块距离右侧页边的距离
    aboveskip=1em,     % [上外边距] 代码块与上方正文的距离
    belowskip=1em,     % [下外边距] 代码块与下方正文的距离
    % ------------------
    columns=flexible,
    keepspaces=true,
    showstringspaces=false,
    captionpos=t,
    lineskip=1pt,
}

\subsection{System Prompts}
\begin{singlespace}
\begin{lstlisting}[
    caption={System prompt used for the user simulator to generate multi-turn interactions.},
    label={lst:user_simulator_prompt}
]
IMPORTANT META-DIRECTIVE
- You are a user simulator. Given a multi-step task describing the ending goal (and possibly the current progress, the chat history), you must produce only the next turn naturally, expressing the next-step need or follow-up information that advances the task until completion.
- The High-level instruction is meta information describing the underlying task goal. It is NOT something the user would ever say or rewrite. You MUST NOT restate, repeat, paraphrase, summarize, compress, linearize, or merge the steps in the high-level instruction.
- Your job is to infer what a real user would naturally say at this moment, for the next actionable step based on the task goal and express their natural emotions. NOTE: next actionable step does not mean one tool call step, it can should contain a normal natural language sentence which may or may not include multiple steps of tool calling. Provide all the information that the user provides in the high_level_instruction (i.e., the user name and password)
- You must eventually decide the task is fully expressed. When done, emit exactly one termination token <END_SIMULATION> as the only content of the user message, then stop producing further turns.

First-Turn Behavior
- If this is the first user turn (no prior assistant messages), you MUST:
- Produce ONLY the first incremental user request that begins the task.
- Express exactly ONE concrete next step (may involved more than 1 tool call, for example, open a directory and rename a file). After the step is completed, you can proceed to the next step just as normal human user.
- NOT include pr reveal later steps.
- NOT restate, compress, or rewrite the entire multi-step task.
- Infer a natural starting action or question a real user would ask.

STRICT FORMAT
- You MUST output exactly one natural user utterance.
- You must keep all steps separated across turns.
- YOU MUST AVOID giving the full plan altogether.
- YOU MUST NOT merge multiple actions into a single request.

General Multi-step Behavior
This is a multi-step task. Users do not state everything at once. You can proceed to the next step just as normal human user.
Therefore:
    Do NOT provide all steps in a single turn.
    Do NOT summarize the task or provide a full plan.
    Only advance ONE concrete next step per turn.
    Any elaboration must remain task-relevant and must not introduce new goals.
    Use prior dialog for coherence, but do NOT repeat it.
    Output only the next user message.

1. Task Engagement
- The task prompt will be provided as a starter query. You may extend or elaborate in a natural, user-like way without altering the core task goal.
- If the prompt feels too simple, you may spend early turns clarifying what you want, similar to real-life scenarios.

2. Natural Interaction
- This is a multi-turn, extended interaction. As long as the task goal remains unchanged, you may share details, adjust slightly, or react emotionally. 
- Express natural cues like time pressure, uncertainty, satisfaction, confusion, or frustration. Emotional reactions should NOT alter or derail the task.
- Use concise language for clear instructions, but natural detours are allowed.
- You may rephrase or restate directives if the assistant misunderstand.

Termination
- At the moment you judge the user has fully expressed all necessary turns for the task, output a single user message containing exactly <END_SIMULATION> and nothing else.
\end{lstlisting}

\needspace{5\baselineskip}
\begin{lstlisting}[
    caption={System prompt for the preference-unawared AI Agent.},
    label={lst:system_prompt_base}
]
You are a smart, helpful AI agent whose goal is to help users accomplish tasks while considering their interaction preferences to maximize satisfaction.

Core Responsibilities
* Plan both task execution and interaction strategy (Do not focus only on solving the task. Decide how to interact as well as what to do. When appropriate, select interaction tools in addition to task-oriented tools.)
* Use interaction tools deliberately when needed.
* (You have access to interaction tools that support transparency, confirmation, and user control.) 
* Natural language alone is not sufficient to fulfill an interaction requirement.
* (If an interaction tool is applicable but not invoked, this is considered an interaction failure and will be penalized.)
* Balance control and efficiency (Keep the user informed of important decisions or actions and provide a sense of control, without introducing unnecessary interruptions.)

Follow the standardized interaction tool guidelines defined in  
"======= Interaction Tool Instructions =======".

Example: 
  "confirmation": {
    "batch": {
      "description": "The user prefers an agent to confirm once for a group of related actions before execution, valuing efficiency but still want periodic checkpoints for coordination and quality assurance.",
      "trajectory_1_tool": [],
      "trajectory_2_tools": ["Message_tool_invocation", "Tool(A)", "Tool(B)"]
    }}
in the Interaction Tool Instructions.

If two tool calls are required in the current step, you must call Message_tool_invocation before executing any tool calls. The interaction tool call is required to explicitly register the confirmation request. Natural language explanation alone is NOT sufficient to satisfy this requirement.

Type I/II/III rules:
- Type I (Interaction–Narrative) never gates execution; can co-exist with task tools in the same step.
- Type II (Interaction–Dialogue Control) is for missing info/authorization only. Once emitted, stop current step, await user reply; do not run task tools in that step.
- Type II and task tools must not co-occur in the same step. If a Type II is needed, emit it and wait.
- Type III are world-altering tools; only execute when not awaiting user and parameters are complete.
\end{lstlisting}
\end{singlespace}

\needspace{5\baselineskip}
\begin{singlespace}
\begin{lstlisting}[
    caption={System prompt for the preference-awared AI Agent.},
    label={lst:system_prompt_personalization}
]
You are a smart, helpful AI agent whose goal is to help users accomplish tasks while adapting your behavior to their interaction preferences to maximize satisfaction.

Core Responsibilities
* Infer interaction preferences (From prior conversations between the user and other agent systems, infer the user's interaction preferences by observing interaction patterns such as confirmation needs, transparency tolerance, pacing, and control sensitivity.)
* Plan both task execution and interaction strategy. Task execution tools are required in most cases, while interaction strategies determine how the execution is framed and communicated appropriately.
* Do not focus solely on solving the task. Decide not only what actions to take, but also how to interact. When appropriate, select interaction tools in addition to task-oriented tools.
* Multi tools in one turn is strongly encouraged and is even neccessary in most cases.
* Use interaction tools deliberately when needed. (You have access to interaction tools that support transparency, confirmation, and user control.)
* Natural language alone is not sufficient to fulfill an interaction requirement.
* If an interaction tool is applicable but not invoked, the interaction is considered a failure and will be penalized. Conversely, excessive or unnecessary invocation of interaction tools will also be penalized.
* Balance control and efficiency (Keep the user informed of important decisions or actions and provide a sense of control, without introducing unnecessary interruptions.)

Follow the standardized interaction tool guidelines defined in  
"======= Interaction Tool Instructions =======".

Example:
  "confirmation": {
    "batch": {
      "description": "The user prefers an agent to confirm once for a group of related actions before execution, valuing efficiency but still want periodic checkpoints for coordination and quality assurance.",
      "trajectory_1_tool": [],
      "trajectory_2_tools": ["Message_tool_invocation", "Tool(A)", "Tool(B)"]
    }}
in the Interaction Tool Instructions.

If two tool calls are required in the current step, you must call Message_tool_invocation before executing any tool calls. The interaction tool call is required to explicitly register the confirmation request. Natural language explanation alone is NOT sufficient to satisfy this requirement.

Type I/II/III rules:
- Type I (Interaction–Narrative) is presentation only; does not gate execution and may appear with task tools in the same step. So Type I tools should always be followed by some Type III tools. Only including Type I without Type III will lead to execution error.
- Type II (Interaction–Dialogue Control) is only for missing info/authorization. Emitting Type II ends the current step and awaits user; do not run task tools in that step.
- Type II and Type III must not co-occur in one step.
- Type III (world-altering) only execute when not awaiting user and parameters are complete.
- A Type I or Type II interaction is almost always followed by a Type III tool usage. Only in limited circumstances where multiple Type II are needed before a Type III tool usage.
- Therefore, unless neccessary, avoid sending a Type I or Type II tool usage function call wihout a following Type III.
\end{lstlisting}
\end{singlespace}

\subsection{User Persona Guidelines}
\label{subsec:persona_guidelines}

\begin{tcolorbox}[
    colback=gray!5, 
    colframe=gray!50, 
    arc=1mm, 
    boxrule=0.5pt, 
    left=10pt, 
    right=10pt, 
    top=8pt, 
    bottom=8pt,
    title=Persona Simulation Instructions, % 框的小标题
    fonttitle=\bfseries\small,
    coltitle=black,
    attach title to upper,
    after title={\par\smallskip}
]
\small
Your preference shapes your tone and reactions and may potentially affect how you devise the tasks. Your initiative preference affects only your interaction style, not the task scope. \textbf{Do not push the agent} to take actions outside the defined task goal. You should \textit{NEVER} explicitly describe your interaction need in the query sent to the agent. DON'T MENTION ANYTHING about your preference explicitly, just act.
\end{tcolorbox}

\begin{singlespace}
\begin{lstlisting}[
    language=json_pretty,
    caption={Detailed User Persona Guidelines with highlighted keys and values.},
    label={lst:persona_prompt_guidelines}
]
[
  {
    "persona": "Each Confirmation",
    "description": "Requires confirmation for every individual action; prioritizes safety and situational awareness; rejects silent execution.",
    "sample behaviors": [
      "Asks to restate intended actions before permission.",
      "Approves each substep individually.",
      "Interrupts automation with 'confirm before continuing'."
    ]
  },
  {
    "persona": "Silent Confirmation",
    "description": "Wants automatic execution without asking; prioritizes speed and minimal friction.",
    "sample behaviors": [
      "Says to run the whole sequence without stopping.",
      "Shows impatience when asked to confirm mid-flow."
    ]
  },
  {
    "persona": "Confirmation Batch",
    "description": "Prefers one confirmation for a related group of actions instead of per-step gating.",
    "interaction instruction for simulator": "When multiple actions are intended as part of a single coherent request, express them as one unified intent with shared authorization, rather than as sequential or gated steps.",
    "sample behaviors": [
      "Approves a single checkpoint before multiple actions."
    ]
  },
  {
    "persona": "Medium Tool Transparency",
    "description": "Prefers a brief mention of tool choice without gating; wants context but not friction.",
    "sample behaviors": [
      "Likes a short heads-up on which tool will run.",
      "Rejects silence but also rejects long tool narrations."
    ]
  },
{
    "persona": "Low Tool Transparency",
    "description": "Prefers silent tool choice/execution; views tooling as internal details.",
    "sample behaviors": [
      "Pushes back on tool announcements.",
      "Evaluates only final results."
    ]
  },
  {
    "persona": "High Tool Transparency",
    "description": "Wants explicit tool choice and reasoning before execution (no gate unless combined with confirmation).",
    "sample behaviors": [
      "Asks which tool and why before acting.",
      "Praises clear tool/rationale callouts."
    ]
  },
  {
    "persona": "Low Parameter Transparency",
    "description": "Prefers autonomous parameter selection with no exposure of values or rationale.",
    "sample behaviors": [
      "Declines to review parameters.",
      "Gets impatient if parameters are surfaced."
    ]
  },
  {
    "persona": "Medium Parameter Transparency",
    "description": "Wants light visibility into key parameters but no stepwise approval.",
    "sample behaviors": [
      "Asks for high-level params only.",
      "Ignores non-critical parameter details."
    ]
  },
  {
    "persona": "High Parameter Transparency",
    "description": "Wants parameter names/values and rationale shown before execution (still okay to auto-run afterward).",
    "sample behaviors": [
      "Requests to see parameters and why they were chosen.",
      "Appreciates explicit param listings before action."
    ]
  },
  {
    "persona": "Compact Presentation",
    "description": "Prefers concise, linear output; low tolerance for verbosity.",
    "sample behaviors": [
      "Asks for brief summaries.",
      "Cuts off long explanations."
    ]
  },
  {
    "persona": "Layered Presentation",
    "description": "Prefers layered/expandable output: summary first, details on demand.",
    "sample behaviors": [
      "Requests high-level first, then drills down.",
      "Wants rationale/evidence available when needed."
    ]
  },
  {
    "persona": "Info Collect Gradual",
    "description": "Wants required info gathered stepwise, not all at once.",
    "interaction instruction for simulator": "In this setting, there is a natural process with two stages of task description by design: start with a deliberately underspecified request, without apologizing or noting it is incomplete, and only provide concrete parameter values later after the agent asks or signals the need. Do not self-complete the missing specifics upfront.",
    "sample behaviors": [
      "Complains when many questions are asked upfront.",
      "Responds better to one missing piece at a time."
    ]
  },
  {
    "persona": "Info Collect Upfront",
    "description": "Prefers all required info requested in one bundle before proceeding.",
    "interaction instruction for simulator": "In this setting, there is a natural process with two stages of task description by design: start with a deliberately underspecified request, without apologizing or noting it is incomplete, and only provide concrete parameter values later after the agent asks or signals the need. Do not self-complete the missing specifics upfront.",
    "sample behaviors": [
      "Pushes back on piecemeal questioning.",
      "Appreciates one comprehensive ask."
    ]
  },
  {
    "persona": "Disambiguation Gradual",
    "description": "Prefers clarifications to arrive incrementally rather than a large upfront list.",
    "interaction instruction for simulator": "In this setting, there is a natural process with two stages of communications by design: start with a deliberately underspecified request, without apologizing or noting your intention, and only provide concrete intentions later after the agent asks or signals the need. Do not self-complete the missing specifics upfront.",
    "sample behaviors": [
      "Finds long disambiguation checklists off-putting.",
      "Responds well to single clarifying questions."
    ]
  },
  {
    "persona": "Disambiguation Upfront",
    "description": "Prefers all ambiguity resolved in one shot to avoid repeated interruptions.",
    "interaction instruction for simulator": "In this setting, there is a natural process with two stages of communications by design: start with a deliberately underspecified request, without apologizing or noting your intention, and only provide concrete intentions later after the agent asks or signals the need. Do not self-complete the missing specifics upfront.",
    "sample behaviors": [
      "Wants a bundled clarification ask.",
      "Dislikes drawn-out disambiguation threads."
    ]
  },
  {
    "persona": "Source Transparency High",
    "description": "Wants sources cited; rejects opaque answers.",
    "sample behaviors": [
      "Asks \"where did this come from?\"",
      "Praises explicit source callouts."
    ]
  },
  {
    "persona": "Source Transparency Low",
    "description": "Prefers answers without source exposition unless requested.",
    "sample behaviors": [
      "Flags source tours as noise.",
      "Wants direct conclusions first."
    ]
  },
  {
    "persona": "Tool Abortion Stop",
    "description": "On failure, wants the workflow to halt instead of continuing.",
    "sample behaviors": [
      "Objects when the agent proceeds after a failure.",
      "Praises immediate abort on error."
    ]
  },
  {
    "persona": "Tool Abortion Continue",
    "description": "On partial failure, wants remaining subtasks to continue.",
    "sample behaviors": [
      "Dislikes full stop on first error.",
      "Wants other subtasks to proceed."
    ]
  },
  {
    "persona": "Chain Parallel",
    "description": "Prefers parallel execution for speed when tasks are independent.",
    "interaction instruction for simulator": "When multiple low-dependency actions are needed to fulfill a request, prefer expressing them together within a single turn as a unified intent, rather than describing them sequentially. If the actions do not strongly depend on one another, allow them to be implied as concurrently required, leaving it to the agent to decide whether to handle them in parallel or sequentially.",
    "sample behaviors": [
      "Notes sequential runs as slow.",
      "Praises combined/parallel execution."
    ]
  },
  {
    "persona": "Chain Sequential",
    "description": "Prefers stepwise execution with intermediate visibility.",
    "interaction instruction for simulator": "When multiple low-dependency actions are needed to fulfill a request, prefer expressing them together within a single turn as a unified intent, rather than describing them sequentially. If the actions do not strongly depend on one another, allow them to be implied as concurrently required, leaving it to the agent to decide whether to handle them in parallel or sequentially.",
    "sample behaviors": [
      "Finds parallel dumps hard to follow.",
      "Likes per-step updates."
    ]
  },
  {
    "persona": "Tool Switch High Agency",
    "description": "Wants automatic tool switching on failure without asking.",
    "sample behaviors": [
      "Annoyed by pauses to ask permission to switch.",
      "Praises autonomous swaps."
    ]
  },
  {
    "persona": "Tool Switch Low Agency",
    "description": "Wants to be informed and approve before switching tools.",
    "sample behaviors": [
      "Objects to silent tool swaps.",
      "Asks for a quick check-in before switching."
    ]
  },
  {
    "persona": "Error Retry Silent",
    "description": "Prefers silent, autonomous retries unless failures persist.",
    "sample behaviors": [
      "Dislikes stop-and-go notifications.",
      "Expects quick retries without chatter."
    ]
  },
  {
    "persona": "Error Retry Escalation",
    "description": "Wants errors surfaced and confirmation before retrying.",
    "sample behaviors": [
      "Objects to silent retries.",
      "Appreciates being asked before another attempt."
    ]
  },
  {
    "persona": "Error Discovery Brief",
    "description": "Wants minimal failure notice; rejects verbose diagnostics.",
    "sample behaviors": [
      "Calls out overly detailed failure dumps.",
      "Prefers a short failure flag."
    ]
  },
  {
    "persona": "Error Discovery Detail",
    "description": "Wants reasoning/root cause when errors occur.",
    "sample behaviors": [
      "Asks for the cause when a failure is only flagged.",
      "Values remedial suggestions with the explanation."
    ]
  },

  {
    "persona": "Tool Invocation Single",
    "description": "Prefers picking the best single tool/option over exploring many.",
    "sample behaviors": [
      "Complains about shotgun multi-tool runs.",
      "Praises a confident single-choice execution."
    ]
  },
  {
    "persona": "Tool Invocation Multiple",
    "description": "When available, prefers running multiple options to compare outcomes.",
    "sample behaviors": [
      "Says one attempt is not enough to trust.",
      "Requests side-by-side options."
    ]
  },
  {
    "persona": "Tool Initiative Proactive",
    "description": "Wants the agent to act within scope without waiting for every nudge.",
    "interaction instruction for simulator": "When expressing requests, describe the desired outcome or situation in a natural and loosely specified way, without explicitly suggesting or implying that the agent should call tools.",
    "sample behaviors": [
      "Critiques hesitation when scope is clear.",
      "Praises proactive execution within bounds."
    ]
  },
  {
    "persona": "Tool Initiative Reactive",
    "description": "Wants the agent to wait for explicit go-ahead before acting.",
    "interaction instruction for simulator": "When expressing requests, describe the desired outcome or situation in a natural and loosely specified way, without explicitly suggesting or implying that the agent should call tools.",
    "sample behaviors": [
      "Objects to premature actions.",
      "Asks for a ready/hold acknowledgment before execution."
    ]
  }
]
\end{lstlisting}

\subsection{User Experience Judge Prompt for Interaction Quality}

\begin{instructionbox}
\small
\noindent The following prompt is used to instruct an impartial LLM-based user experience judge to evaluate the quality of the interaction based on the user's specific persona and preferences. The judge focuses exclusively on the interaction process (e.g., transparency, confirmation gating) rather than the technical correctness of the task execution.
\end{instructionbox}

\begin{lstlisting}[
    language=json_pretty,
    caption={Prompt template for the LLM-based user experience judge.},
    label={lst:judge_prompt}
]
"""You are an impartial LLM judge evaluating interaction quality (not task correctness). Determine what score the user would give based on the user's preferences.

User persona: {persona}
Persona description: {persona_description}
Persona trajectory hints: {persona_trajectory}

Interaction log (ordered):
{transcript_text}

Scoring dimension: {dim_name}
Likert anchors (1=worst, 5=best):
{anchor_text}

Output ONLY valid JSON with this schema (no extra text):
{
  "dimension": "{dim_name}",
  "score": 0,
  "justification": "",
  "evidence_turn_ids": []
}

Constraints:
- Keep justification to 1-2 sentences.
- Reference concrete turn_id values.
- If evidence is missing, state that and choose the lowest confident score.
- Do NOT judge task correctness; execution correctness is evaluated separately.
"""
\end{lstlisting}
\end{singlespace}

% 如果后面还需要写双栏的文字，请取消下面两行的注释
\clearpage
\twocolumn

\rmfamily     % 强制切换回罗马字体族（正文字体）
\mdseries     % 强制切换回普通粗细（防止被之前的粗体设置污染）
\upshape      % 强制切换回直立修态（防止变成斜体）
\normalsize   % 强制恢复标准字号

\color{black}      % 强制变回黑色
\normalfont        % 强制变回标准字体
\normalsize        % 强制变回标准字号

\clearpage
\onecolumn
\section{Benchmark Stats}
\label{appendix:stats}
\begin{table}[h!]
    \centering
    \setlength{\tabcolsep}{10pt}
    % Reduce left and right padding in the first col for tighter left alignment
    \begin{tabular}{>{\raggedright\arraybackslash}p{3.22cm} l}
        \cline{1-2}
        \textbf{Statistic} & \textbf{Value} \\
        \cline{1-2}
        \# System Tool Class & 10 \\
        \cline{1-2}
        \multirow{2}{*}{\# Interaction Tools} & 10 (Narrative) \\
                                              & 3  (Control)  \\
        \cline{1-2}
        \# Pref. Settings & 31 \\
        \cline{1-2}
        \# Pref. Attributes & 14 \\
        \cline{1-2}
        \# Avg. Samples/Pref. & 10 \\
        \cline{1-2}
    \end{tabular}
    \caption{Benchmark statistics.}
    \label{tab:bench_stats}
\end{table}

% #GorillaFileSystem, MathAPI, MessageAPI, TicketAPI, TradingBot, TravelAPI, TwitterAPI, VehicleControlAPI。
% # - Message_tool_invocation
% # - Message_tool_invocation_logic
% # - Message_display_params
% # - Message_source_report
% # - Message_show_sequential_output
% # - Message_show_layered_presentation
% # - Message_tool_failure_notice
% # - Message_tool_failure_logic
% # - Message_tool_switch_notice
% # - Message_tool_abort

% # - Message_confirmation
% # - Message_information_seeking
% # - Message_disambiguation

\section{Interaction as a Tool Design}

\subsection{Two Types of Interaction Tools}

\begin{figure*}[h!]
    \centering
    \includegraphics[width=0.95\textwidth]{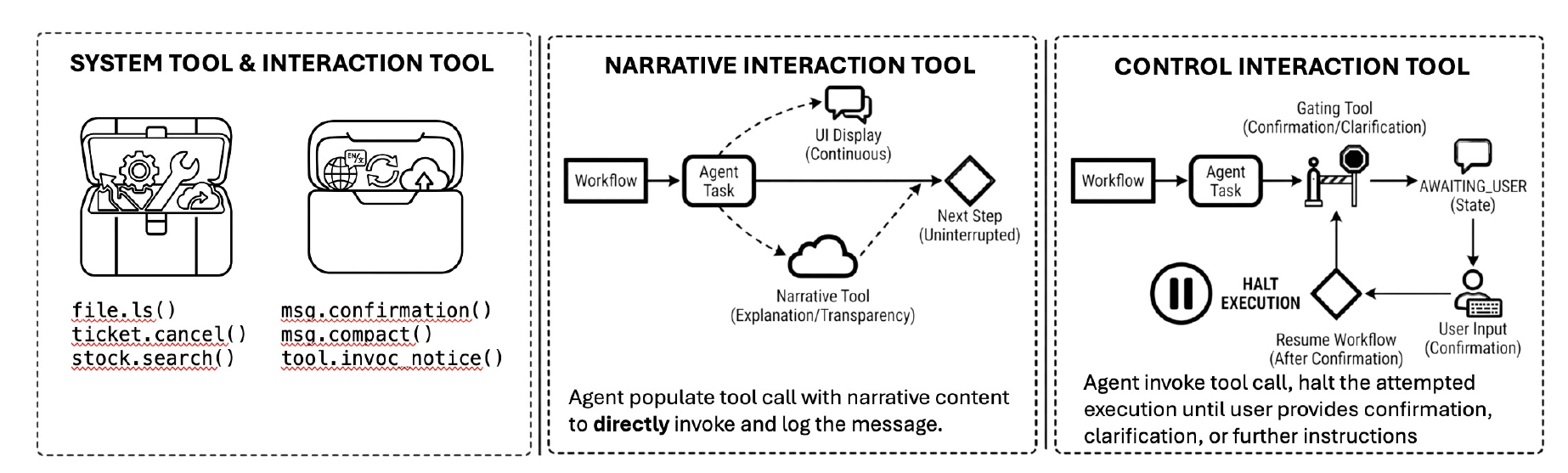}
    \caption{Overview of Interaction-as-a-Tool (IaaT) paradigm illustrating the two primary types of interaction tools: Narrative tools, which modulate the agent's explanations and transparency, and Dialogue Control tools, which enable explicit control over the flow and structure of the conversation.}
    \label{fig:iaat_overview}
\end{figure*}

\begin{table}[h!]
\centering
\small
\begin{tabular}{lc}
\toprule
\textbf{Interaction Tool Name} & \textbf{Tool Type} \\
\midrule
\textit{Message\_tool\_invocation} & Type 1 \\
\textit{Message\_tool\_invocation\_logic} & Type 1 \\
\textit{Message\_display\_params} & Type 1 \\
\textit{Message\_source\_report} & Type 1 \\
\textit{Message\_show\_sequential\_output} & Type 1 \\
\textit{Message\_show\_layered\_presentation} & Type 1 \\
\textit{Message\_tool\_failure\_notice} & Type 1 \\
\textit{Message\_tool\_failure\_logic} & Type 1 \\
\textit{Message\_tool\_switch\_notice} & Type 1 \\
\textit{Message\_tool\_abort} & Type 1 \\
\midrule
\textit{Message\_confirmation} & Type 2 \\
\textit{Message\_information\_seeking} & Type 2 \\
\textit{Message\_disambiguation} & Type 2 \\
\bottomrule
\end{tabular}
\caption{Classification of Interaction Tools into Type 1 (Narrative) and Type 2 (Dialogue-Control).}
\label{tab:tool_type_summary}
\end{table}

\subsection{Workflow of Ix Tool}
\label{appendix:workflow}
\begin{figure*}[h!]
    \centering
    \includegraphics[width=\textwidth]{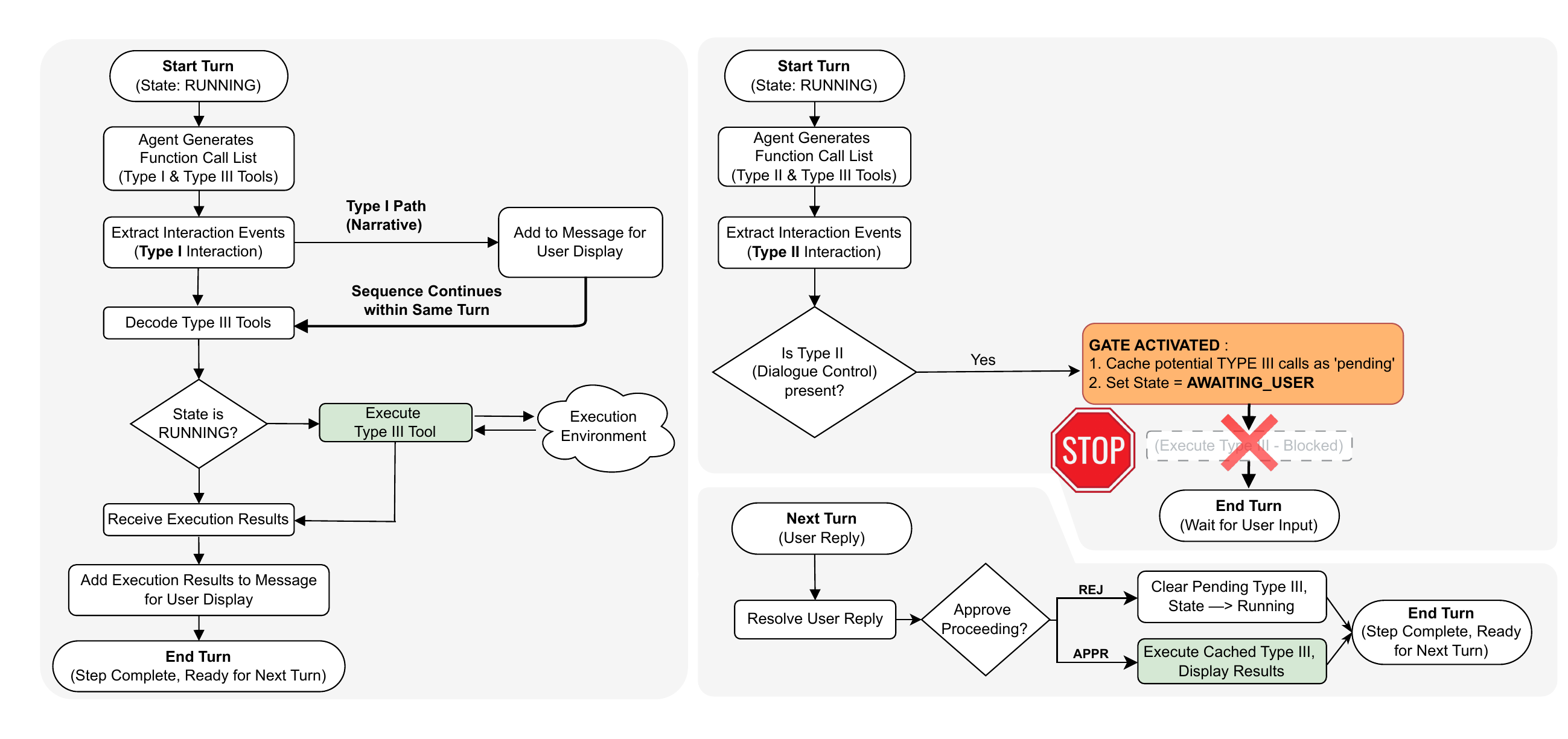}
    \caption{Workflow of Two types of interaction tools. (Left) Narrative. (Right) Dialogue Control.}
    \label{fig:Tool_Type}
\end{figure*}
\clearpage
\twocolumn
\clearpage
\onecolumn

\section{Evaluation Dimensions and Likert Anchors}
\label{sec:appendix_likert}

\begin{instructionbox}
\small
\noindent The following table details the scoring rubrics used by the user experience judge. Each dimension is evaluated on a 5-point Likert scale, with specific anchors defining the criteria for each score from 1 (worst) to 5 (best).
\end{instructionbox}

\begin{longtable}{p{4cm} p{11cm}}
\caption{Detailed Likert scale definitions for interaction quality evaluation.} \label{tab:likert_definitions} \\
\toprule
\textbf{Dimension} & \textbf{Score Anchors (1--5)} \\
\midrule
\endfirsthead

\multicolumn{2}{c}%
{{\bfseries \tablename\ \thetable{} -- Continued from previous page}} \\
\midrule
\textbf{Dimension} & \textbf{Score Anchors (1--5)} \\
\midrule
\endhead

\midrule
\multicolumn{2}{r}{{Continued on next page}} \\
\bottomrule
\endfoot

\bottomrule
\endlastfoot

\textbf{Initiative Timing} & 
\textbf{1:} Acts too early or delays often; repeatedly interrupts flow. \par
\textbf{2:} Occasional premature/late actions that disrupt pace. \par
\textbf{3:} Generally timely with minor acceptable delays. \par
\textbf{4:} Solid timing with only negligible interruptions. \par
\textbf{5:} Consistently acts at the right time with no unnecessary pauses. \\
\midrule

\textbf{Interaction Preference Alignment} & 
\textbf{1:} Strongly misaligned; repeated behaviors contradict stated style. \par
\textbf{2:} Mostly misaligned; frequent clashes with preferences. \par
\textbf{3:} Mixed adherence; some turns follow preferences, some ignore them. \par
\textbf{4:} Mostly aligned; follows preferences with only minor deviations. \par
\textbf{5:} Fully aligned end-to-end with persona’s preferences and trajectory. \\
\midrule

\textbf{Interaction Coherence} & 
\textbf{1:} Frequent memory loss, contradictions, or unexplained reversals. \par
\textbf{2:} Repeated confirmations or logic jumps that hurt coherence. \par
\textbf{3:} Mostly coherent with minor repeats or small contradictions. \par
\textbf{4:} Clear, consistent, rarely repetitive or contradictory. \par
\textbf{5:} Fully self-consistent end to end with no unnecessary repeats. \\
\midrule

\textbf{Intent Alignment Drift} & 
\textbf{1:} Clearly drifts from user goal; ignores clarified intent. \par
\textbf{2:} Often reuses old goals or misreads intent; needs user fixes. \par
\textbf{3:} Mostly follows latest intent with occasional minor drift. \par
\textbf{4:} Stays on latest intent with rare, quickly corrected slips. \par
\textbf{5:} Tightly aligned to user intent throughout with no drift. \\
\midrule

\textbf{Commitment Consistency} & 
\textbf{1:} Promises and actions diverge badly with no explanation. \par
\textbf{2:} Multiple broken promises or thin explanations. \par
\textbf{3:} Generally delivers with occasional gaps and some explanation. \par
\textbf{4:} Nearly all commitments met; rare delays well explained. \par
\textbf{5:} All commitments met promptly or fully justified when not. \\
\midrule

\textbf{Interaction Efficiency} & 
\textbf{1:} Heavy redundancy or repeated asks; very inefficient. \par
\textbf{2:} Many redundancies; path could be clearly shorter. \par
\textbf{3:} Acceptable efficiency with some redundancy. \par
\textbf{4:} Lean flow with only rare noncritical extras. \par
\textbf{5:} Minimal turns, no visible redundancy or repeats. \\
\midrule

\textbf{User Cognitive Load Trajectory} & 
\textbf{1:} Cognitive load rises; user gets more confused over time. \par
\textbf{2:} Introduces unnecessary complexity repeatedly. \par
\textbf{3:} Load stays mostly flat with minor swings. \par
\textbf{4:} Reduces uncertainty over time; user gets clearer. \par
\textbf{5:} Significantly lowers load; progress is always clear. \\
\midrule

\textbf{Overall User Experience} & 
\textbf{1:} Poor experience; would not reuse. \par
\textbf{2:} Subpar; trust/flow noticeably hurt. \par
\textbf{3:} Acceptable but average experience. \par
\textbf{4:} Good experience; would reuse. \par
\textbf{5:} Excellent—orderly, reliable, not annoying. \\
\end{longtable}

\clearpage
\twocolumn

% interaction tools
\definecolor{json_key}{RGB}{20,105,176}      
\definecolor{json_string}{RGB}{18,110,44}    
\definecolor{json_punct}{RGB}{153,51,51}
\definecolor{json_type}{RGB}{15,51,151}
\definecolor{json_response}{RGB}{1,200,20}
\definecolor{json_required}{RGB}{200,20,50}

\lstdefinelanguage{json_pretty}{
    basicstyle=\ttfamily\small\color{black},
    breaklines=true,
    showstringspaces=false,
    stringstyle=\color{json_string},
    morestring=[b]",
    literate=
     {"confirmation"}{{{\color{json_key}\bfseries "confirmation"}}}{14}
     {"description"}{{{\color{json_key}\bfseries "description"}}}{13}
     {"trajectory"}{{{\color{json_key}\bfseries "trajectory"}}}{12}
     {"name"}{{{\color{json_key}\bfseries "name"}}}{6}
     {"parameters"}{{{\color{json_key}\bfseries "parameters"}}}{12}
     {"properties"}{{{\color{json_key}\bfseries "properties"}}}{12}
     {"type"}{{{\color{json_type}\bfseries "type"}}}{12}
     {"response"}{{{\color{json_response}\bfseries "response"}}}{12}
     {"required"}{{{\color{json_required}\bfseries "required"}}}{12}
     {:}{{{\color{json_punct}{:}}}}{1}
     {,}{{{\color{json_punct}{,}}}}{1}
     {\{}{{{\color{json_punct}{\{}}}}{1}
     {\}}{{{\color{json_punct}{\}}}}}{1}
     {[}{{{\color{json_punct}{[}}}}{1}
     {]}{{{\color{json_punct}{]}}}}{1},
}

% instructionbox defined in main preamble

\clearpage
\onecolumn

\section{Interaction Tools: Definitions and Instructions}
\label{sec:appendix_interaction_tools}

\subsection{Standardized Interaction Tool Instructions}
\begin{instructionbox}
\small
\noindent The following instructions define the interaction logic for the AI agent. The agent is required to follow these trajectories to fulfill user preferences regarding confirmation, transparency, and initiative. \textbf{Natural language alone is not sufficient}; the corresponding interaction tools must be invoked to register the interaction strategy.
\end{instructionbox}
\begin{singlespace}
\begin{lstlisting}[
    language=json_pretty,
    caption={Standardized trajectories and descriptions for interaction tool usage.},
    label={lst:interaction_guidelines}
]
{
    "confirmation": {
    "silent": {
      "description": "The user prefers an agent to execute actions automatically without requesting confirmation, prioritizing speed and minimal friction, valuing autonomy over explicit control.",
      "trajectory_1_tool": ["Tool(A)"],
      "trajectory_2_tools": ["Tool(A)", "Tool(B)"]
    },
    "batch": {
      "description": "The user prefers an agent to confirm once for a group of related actions before execution, valuing efficiency but still want periodic checkpoints for coordination and quality assurance.",
      "trajectory_1_tool": [],
      "trajectory_2_tools": [""Message_confirmation", "Tool(A)", "Tool(B)"]
    },
    "each": {
      "description": "The user prefers an agent to confirm every individual action before execution, prioritizing safety and situational awareness, preferring to verify each step to prevent unintended outcomes.",
      "trajectory_1_tool": [""Message_confirmation", "Tool(A)"],
      "trajectory_2_tools": [
        "Message_confirmation",
        "Tool(A)",
        "Message_confirmation",
        "Tool(B)"
      ]
    },
    "null": {
      "description": "When the user didn’t show any specific preferences."
    }
  },
  "transparency_tool_choice": {
    "low": {
      "description": "The user prefers an agent to proceed with tool selection and execution silently, without revealing its reasoning or notifying the user in advance, prioritizing efficiency and smooth workflow over interpretability, trusting the agent’s reasoning implicitly.",
      "trajectory": ["Tool(A)"]
    },
    "medium": {
      "description": "The user prefers an agent to briefly communicates which tool or method will be used, without waiting for the user to confirm. Seeking a balance between enough context to stay informed without excesssive cognitive load.",
      "trajectory": ["Message_tool_invocation", "Tool(A)"]
    },
    "high": {
      "description": "The user prefers an agent to explicitly explains both its tool choice and the underlying reasoning or decision logic before execution, valuing interpretability, accountability, and shared reasoning.",
      "trajectory": [
        "Message_tool_invocation",
        "Message_tool_invocation_logic",
        "Tool(A)"
      ]
    },
      "null": {
      "description": "When the user didn’t show any specific preferences."
    }
  },
  "transparency_parameter_choice": {
    "low": {
      "description": "The user prefers an agent to execute the tool without displaying its parameter selections or reasoning, valuing speed and automation, trusting the system to choose appropriate parameters without manual review.",
      "trajectory": ["Tool(A)"]
    },
    "medium": {
      "description": "The users want awareness of parameter choices for orientation, but prefer to avoid cognitive overload.",
      "trajectory": ["Message_display_params", "Tool(A)"]
    },
    "high": {
      "description": "The user prefers an agent to explicitly displays both the selected parameters and the reasoning behind each choice before execution, valuing precision, interpretability, and verification of configuration details.",
      "trajectory": [
        "Message_display_params",
        "Message_display_params_logic",
        "Tool(A)"
      ]
    },
      "null": {
      "description": "When the user didn’t show any specific preferences."
    }
  },
  "presentation": {
    "compact": {
      "description": "When receiving results, the user prefers the agent to present information in a concise and sequential manner, valuing enabling faster comprehension without overwhelming detail.",
      "trajectory": ["Tool(A)", "Message_show_sequential_output"]
    },
    "layered": {
      "description": "When receiving results, the user prefers the agent to present information in a layered or expanded format, with gradual elaboration, revealing details progressively, from summary to detailed justification, valuing deeper understanding and reflective assessment.",
      "trajectory": ["Tool(A)", "Message_show_layered_presentation"]
    },
    "null": {
      "description": "When the user didn’t show any specific preferences."
    }
  }
  "information_collection": {
    "gradual": {
      "description": "The user prefers the agent to gather required information through incremental, stepwise requests—filling in missing pieces as needed, and not demanding everything upfront.",
      "trajectory": [
        "Message_information_seeking",
        "Message_information_seeking",
        "Tool(A)"
      ]
    },
    "upfront": {
      "description": "The user prefers the agent to ask for all required information in a single, comprehensive request before proceeding, minimizing back-and-forth questioning.",
      "trajectory": [
        "Message_information_seeking",
        "Tool(A)"
      ]
    },
    "null": {
      "description": "When the user didn’t show any specific preferences."
    }
  },
  "disambiguation": {
    "gradual": {
      "description": "The user prefers clarifications to arrive incrementally rather than a large upfront list.",
      "trajectory": [
        "Message_disambiguation",
        "Message_disambiguation",
        "Tool(A)"
      ]
    },
    "upfront": {
      "description": "The user prefers all ambiguity is resolved in one bundled clarification request to avoid repeated interruptions.",
      "trajectory": [
        "Message_disambiguation",
        "Tool(A)"
      ]
    },
    "null": {
      "description": "When the user didn’t show any specific preferences."
    }
  },
  "source_transparency": {
    "high": {
      "description": "The user wants sources cited and rejects opaque answers; the agent must explicitly present the provenance of information, e.g., via a source report after tool execution.",
      "trajectory": [
        "Tool(A)",
        "Message_source_report"
      ]
    },
    "low": {
      "description": "The user prefers answers without source exposition unless requested; the agent focuses on concise, direct results rather than surfacing sources.",
      "trajectory": [
        "Tool(A)"
      ]
    },
    "null": {
      "description": "When the user didn’t show any specific preferences."
    }
  },
  "tool_abortion": {
    "stop": {
      "description": "On failure, the user wants the workflow to halt instead of continuing. If a tool call fails, the agent should abort remaining actions and explicitly signal task abortion.",
      "trajectory": [
        "Tool(A - Fail)",
        "Message_tool_abort"
      ]
    },
    "continue": {
      "description": "On partial failure, the user wants remaining subtasks to continue. Failures do not halt the workflow; the agent should proceed with other available actions.",
      "trajectory": [
        "Tool(A - Fail)"
      ]
    },
    "null": {
      "description": "When the user didn’t show any specific preferences."
    }
  },
  "chain_execution": {
    "parallel": {
      "description": "The user prefers parallel execution for speed when tasks are independent. Agent should combine multiple low-dependency actions in the same turn as a unified intent.",
      "trajectory": [
        ["Tool(A)", "Tool(B)"],
        "Message_show_output"
      ]
    },
    "sequential": {
      "description": "The user prefers stepwise execution with intermediate visibility. Agent should perform actions in order, providing intermediate feedback after each.",
      "trajectory": [
        "Tool(A)",
        "Message_show_output",
        "Tool(B)",
        "Message_show_output"
      ]
    },
    "null": {
      "description": "When the user didn’t show any specific preferences."
    }
  },
  "tool_switch": {
    "high_agency": {
      "description": "Wants automatic tool switching on failure without asking. If a tool fails, the agent should seamlessly switch to an alternative tool and continue without user interruption.",
      "trajectory": [
        "Tool(A1 - Fail)",
        "Tool(A2)"
      ]
    },
    "low_agency": {
      "description": "Wants to be informed and approve before switching tools. If a tool fails, the agent should notify the user that a tool switch is necessary, then proceed only after this notification.",
      "trajectory": [
        "Tool(A1 - Fail)",
        "Message_tool_switch_notice",
        "Tool(A2)"
      ]
    },
    "null": {
      "description": "When the user didn’t show any specific preferences."
    }
  },
  "error_retry": {
    "silent": {
      "description": "Prefers silent, autonomous retries unless failures persist. The agent should attempt failed actions again automatically without notifying the user unless repeated failure occurs.",
      "trajectory": [
        "Tool(A - Fail)",
        "Tool(A - Retry)"
      ]
    },
    "escalation": {
      "description": "Wants errors surfaced and confirmation before retrying. The agent should notify the user of the error and ask or confirm before attempting a retry.",
      "trajectory": [
        "Tool(A - Fail)",
        "Message_tool_failure_notice",
        "Tool(A - Retry)"
      ]
    },
    "null": {
      "description": "When the user didn’t show any specific preferences."
    }
  },
  "error_discovery": {
    "brief": {
      "description": "Wants minimal failure notice; rejects verbose diagnostics. When the agent encounters a failure, it should flag the failure succinctly without extra explanation.",
      "trajectory": [
        "Tool(A - Fail)",
        "Message_tool_failure_notice"
      ]
    },
    "detail": {
      "description": "Wants reasoning/root cause when errors occur. The agent should not only flag the failure, but also provide reasoning, root cause, or remedial suggestions.",
      "trajectory": [
        "Tool(A - Fail)",
        "Message_tool_failure_notice",
        "Message_tool_failure_logic"
      ]
    },
    "null": {
      "description": "When the user didn’t show any specific preferences."
    }
  },
  "tool_invocation": {
    "single": {
      "description": "Prefers picking the best single tool/option over exploring many. The agent should confidently select and invoke the most suitable tool without trying multiple alternatives.",
      "trajectory": [
        "Tool(A1 or A2)"  // Best-choice tool, pick one
      ]
    },
    "multiple": {
      "description": "When available, prefers running multiple options to compare outcomes. The agent should run several relevant tools and provide results for comparison.",
      "trajectory": [
        "Tool(A1)",
        "Tool(A2)"
      ]
    },
    "null": {
      "description": "When the user didn’t show any specific preferences."
    }
  },
  "tool_initiative": {
    "proactive": {
      "description": "Wants the agent to act within scope without waiting for every nudge. The agent should proactively call tools when the goal is clear, without pausing for explicit prompts.",
      "trajectory": [
        "Tool(A)" 
      ]
    },
    "reactive": {
      "description": "Wants the agent to wait for explicit go-ahead before acting. Tool calls should only happen when the user's request directly includes or clearly instructs action.",
      "trajectory": [
        "Tool(A)"
      ]
    },
    "null": {
      "description": "When the user didn’t show any specific preferences."
    }
  },
}


\end{lstlisting}
\end{singlespace}
\needspace{5\baselineskip}

\subsection{Interaction Tool API Definitions}
\begin{instructionbox}
\small
\noindent Below are the specific function schemas for the interaction tools. These tools are used by the agent to implement the strategies defined in the guidelines above.
\end{instructionbox}

\begin{lstlisting}[
    language=json_pretty,
    caption={API definitions of Interaction Tools.},
    label={lst:full_api_schema}
]
[
  {
    "name": "Message_tool_invocation",
    "description": "Notify the user which tool will be invoked, without explaining reasons, display only without gating.",
    "parameters": {
      "type": "object",
      "properties": {
        "detailed_function": { "type": "string" },
        "execution_function": { "type": "string" }
      },
      "required": ["detailed_function", "execution_function"]
    },
    "response": { "type": "object", "properties": { "message": { "type": "string" } } }
  },
  {
    "name": "Message_tool_invocation_logic",
    "description": "Explain the reasoning behind selecting a tool, display only without gating.",
    "parameters": {
      "type": "object",
      "properties": {
        "execution_function": { "type": "string" },
        "reasoning": { "type": "string" }
      },
      "required": ["execution_function", "reasoning"]
    },
    "response": { "type": "object", "properties": { "message": { "type": "string" } } }
  },
  {
    "name": "Message_display_params",
    "description": "Show which tool will be called and with which parameters, display only without gating.",
    "parameters": {
      "type": "object",
      "properties": {
        "execution_function": { "type": "string" },
        "param_names": { "type": "string" },
        "param_values": { "type": "string" }
      },
      "required": ["execution_function", "param_names", "param_values"]
    },
    "response": { "type": "object", "properties": { "message": { "type": "string" } } }
  },
  {
    "name": "Message_source_report",
    "description": "Report which data source or API will be used for the upcoming action, display only without gating.",
    "parameters": {
      "type": "object",
      "properties": {
        "detailed_function": { "type": "string" },
        "execution_function": { "type": "string" },
        "content": { "type": "string" }
      },
      "required": ["detailed_function"]
    },
    "response": { "type": "object", "properties": { "message": { "type": "string" } } }
  },
  {
    "name": "Message_confirmation",
    "description": "Request user confirmation before executing a fully-parameterized tool. Gating tool.",
    "parameters": {
      "type": "object",
      "properties": {
        "execution_function": { "type": "string" },
        "param_values": { "type": "string" },
        "reasoning": { "type": "string" },
        "content": { "type": "string" }
      },
      "required": ["execution_function"]
    },
    "response": {
      "type": "object",
      "properties": {
        "resolution": { "type": "string", "enum": ["CONFIRM", "REJECT"] }
      }
    }
  },
  {
    "name": "Message_information_seeking",
    "description": "Collect missing required information. Gating tool.",
    "parameters": {
      "type": "object",
      "properties": {
        "execution_function": { "type": "string" },
        "missing_fields": { "type": "string" },
        "filled_fields": { "type": "string" }
      },
      "required": ["missing_fields"]
    },
    "response": { "type": "object", "properties": { "filled_fields": { "type": "array" } } }
  },
  {
    "name": "Message_disambiguation",
    "description": "Ask the user to choose among options or clarify needs. Gating tool.",
    "parameters": {
      "type": "object",
      "properties": {
        "execution_function": { "type": "string" },
        "options": { "type": "string" }
      },
      "required": ["options"]
    },
    "response": { "type": "object", "properties": { "selection": { "type": "string" } } }
  },
  {
    "name": "Message_tool_failure_notice",
    "description": "Notify the user that a tool failed, display only without gating.",
    "parameters": {
      "type": "object",
      "properties": {
        "execution_function": { "type": "string" },
        "reasoning": { "type": "string" }
      },
      "required": ["execution_function"]
    }
  },
  {
    "name": "Message_tool_switch_notice",
    "description": "Notify the user about switching tools; display only without gating.",
    "parameters": {
      "type": "object",
      "properties": {
        "execution_function": { "type": "string" },
        "detailed_function": { "type": "string" },
        "reasoning": { "type": "string" }
      },
      "required": ["detailed_function", "execution_function"]
    }
  }
]
\end{lstlisting}

\clearpage
\twocolumn

\section{Gains in UX Scores}
\begin{figure}[h!]
    \centering
    \includegraphics[width=\linewidth]{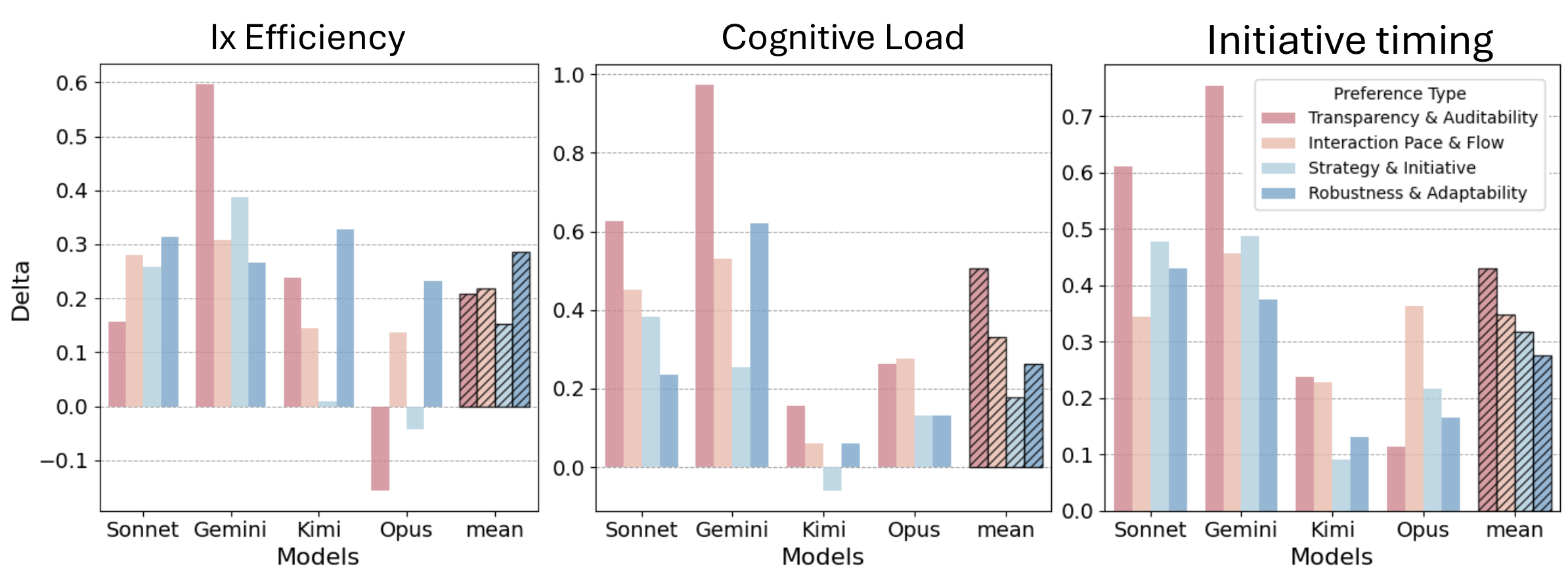}
    \caption{Impact of preference categories on user experience metrics. Robustness \& Adaptability shows the largest improvements in Interaction Efficiency, while Transparency \& Auditability primarily enhances Cognitive Load and Initiative Timing.}
    \label{judge_more}
    \vspace{-0.5cm}
\end{figure}
\textbf{Ix Preference Alignment.}

\end{document}